%Paper: hep-th/9403096
%From: Stefan Theisen <SJT%DMUMPIWH.BITNET@vm.gmd.de>
%Date: Thu, 17 Mar 94 13:00:28 GMT

\input harvmac
\baselineskip=12pt
\def\sst{\scriptscriptstyle}
\def\frac#1#2{{#1\over#2}}

\def\journal#1&#2(#3){\unskip, #1~\bf #2 \rm(19#3) }
\def\andjournal#1&#2(#3){\sl #1~\bf #2 \rm (19#3) }

\def\bra#1{\left\langle #1\right|}
\def\ket#1{\left| #1\right\rangle}
\def\det{{\rm det}}

\catcode`\@=11\def\slash#1{\mathord{\mathpalette\c@ncel{#1}}}
\overfullrule=0pt
\def\steepslash{\c@ncel}
\def\frac#1#2{{#1\over #2}}
\def\l{\lambda}
\def\e{\epsilon}
\def\hb{\hbar}
\def\p{\partial}
\def\:{\!:\!}
\def\inbar{\,\vrule height1.5ex width.4pt depth0pt}
\def\IQ{\relax\,\hbox{$\inbar\kern-.3em{\rm Q}$}}
\def\IB{\relax{\rm I\kern-.18em B}}
\def\IC{\relax\hbox{$\inbar\kern-.3em{\rm C}$}}
\def\IP{\relax{\rm I\kern-.18em P}}
\def\IR{\relax{\rm I\kern-.18em R}}
\def\ZZ{\relax\ifmmode\mathchoice
{\hbox{Z\kern-.4em Z}}{\hbox{Z\kern-.4em Z}}
{\lower.9pt\hbox{Z\kern-.4em Z}}
{\lower1.2pt\hbox{Z\kern-.4em Z}}\else{Z\kern-.4em Z}\fi}

\catcode`\@=12

%                      Zeitschriften:
\def\npb#1(#2)#3{{ Nucl. Phys. }{B#1} (#2) #3}
\def\plb#1(#2)#3{{ Phys. Lett. }{#1B} (#2) #3}
\def\pla#1(#2)#3{{ Phys. Lett. }{#1A} (#2) #3}
\def\prl#1(#2)#3{{ Phys. Rev. Lett. }{#1} (#2) #3}
\def\mpla#1(#2)#3{{ Mod. Phys. Lett. }{A#1} (#2) #3}
\def\ijmpa#1(#2)#3{{ Int. J. Mod. Phys. }{A#1} (#2) #3}
\def\cmp#1(#2)#3{{ Comm. Math. Phys. }{#1} (#2) #3}
\def\cqg#1(#2)#3{{ Class. Quantum Grav. }{#1} (#2) #3}
\def\jmp#1(#2)#3{{ J. Math. Phys. }{#1} (#2) #3}
\def\anp#1(#2)#3{{ Ann. Phys. }{#1} (#2) #3}
\def\prd#1(#2)#3{{ Phys. Rev. } {D{#1}} (#2) #3}
\def\ptp#1(#2)#3{{ Progr. Theor. Phys. }{#1} (#2) #3}
\def\aom#1(#2)#3{{ Ann. Math. }{#1} (#2) #3}

\def\sst{\scriptscriptstyle}
\def\st{\seveni}

\def\a{\alpha}
\def\b{\beta}

\def\g{\gamma}

\def\l{\lambda}
\def\e{\epsilon}
\def\t{\theta}
\def\T{\Theta}
\def\cD{{\cal D}}

\def\{{\lbrace}
\def\}{\rbrace}
\def\({\lbrack}
\def\){\rbrack}

\def\tu{\Theta_u}
\def\tv{\Theta_v}

\def\hb{\hfill\break}
\def\bra{\langle}
\def\ket{\rangle}
\def\pd{\partial}

\def\Z{{\bf Z}}
\def\P{{\bf P}}
\def\pp#1{{\partial\over\partial\rho_{#1}}}

\def\hb{\hfill\break}
\overfullrule=0pt
\def\hb{\hfill\break\noindent}

\def\cicy#1(#2|#3)#4{\left(\matrix{#2}\right|\!\!
                     \left|\matrix{#3}\right)^{{#4}}_{#1}}

%%%%%%%%%%%%%%%%%%%%Hosono macros%%%%%%%%%%%%%%%%%%%

\def\sst{\scriptscriptstyle}
\def\st{\seveni}

\def\a{\alpha}
\def\b{\beta}

\def\g{\gamma}

\def\l{\lambda}
\def\e{\epsilon}
\def\t{\theta}

\def\{{\lbrace}
\def\}{\rbrace}

\def\hb{\hfill\break}
\def\bra{\langle}
\def\ket{\rangle}
\def\U{{\cal U}_\theta}
\def\Ui{\U^{-1}}

\footline{\hss\tenrm--\folio\--\hss}
\nopagenumbers
\rightline{HUTMP-94/01}
\rightline{LMU-TPW-94-02}
\vskip.8cm
\centerline{\titlefont{Lectures on Mirror Symmetry}}\footnote{}
{Extended version of a lecture given at the Third Baltic Student
Seminar, Helsinki 1993}
\vskip.8cm
\centerline{S. Hosono$^{\dagger}$, A. Klemm$^{*}$ and S.
Theisen$^{\diamond}$}
\bigskip
\centerline{$^\dagger$Department of Mathemathics, Toyama University}
\centerline{Toyama 930, Japan}
\medskip
\centerline{$^*$Department of Mathematics, Harvard University}
\centerline{ Cambridge, MA 02138, U.S.A. }
\medskip\centerline{$^{\diamond}$Sektion Physik der Universit\"at
M\"unchen}
\centerline{Theresienstra\ss e 37, D - 80333 M\"unchen, FRG}
\vskip .5in

\noindent
We give an introduction to
mirror symmetry of strings on Calabi-Yau manifolds with an emphasis on
its applications e.g. for the computation
of Yukawa couplings. We introduce all necessary concepts and tools
such as the basics of toric geometry, resolution of singularities,
construction of mirror pairs, Picard-Fuchs equations, etc. and
illustrate all of this on a non-trivial example.
\vskip1cm
\noindent
{\bf Contents:}\hb
\vskip.3cm
\item{1.} Introduction
\item{2.} Strings on Calabi-Yau Manifolds
\item{3.} Superconformal Field Theory and CY Compactification
\item{4.} Construction of Mirror Pairs
\item{5.} Periods, Picard-Fuchs Equations and Yukawa Couplings
\item{6.} Mirror Map and Applications of Mirror Symmetry
\item{7.} A Two Moduli Example: Hypersurface in $\IP^4\(2,2,2,1,1\)$
\item{8.} Conclusions and Outlook

\Date{February 1994}

\vfill\eject

%%%%%%%%%%%%%%%%%%%%%%%%%%%%%%%%%%%%%%%%%%%%%%%%%%%%%%%%%%%%%%
\newsec{Introduction}
%%%%%%%%%%%%%%%%%%%%%%%%%%%%%%%%%%%%%%%%%%%%%%%%%%%%%%%%%%%%%%

Almost ten years after the revival of string theory we are
still a far cry away from being able to convince the critics
of the viability of string theory as a unified description
of elementary particles and their interactions, {\it including}
gravity.

A lot of work has been devoted to the construction of consistent
classical
string vacua and to the study of some of their features, such
as their massless spectrum (i.e. gauge group, number of massless
generations, their representations with respect to the gauge
group, their interactions,
etc.). This line of research is still continuing but
a complete classification seems to be out of reach.

One particular class of string vacua which has received much attention are
compactifications on so-called Calabi-Yau (CY) manifolds.
These notes will deal with some aspects of strings on CY manifolds.
Our goal is to enable the reader to compute the phenomenologically
relevant Yukawa couplings between charged matter fields, which
in a realistic model will have to be identified with quarks,
leptons and Higgs fields. Even though to date there does not
exist a CY model with all the qualitative
features of the standard model,
it is still useful to develop techniques to do explicit computations
in the general case. This is where mirror symmetry enters the stage
in that it allows for the (explicit) computation of a class
of Yukawa couplings which are otherwise very hard (if not impossible)
to obtain.

As it is the purpose of these notes to give a pedagogical
introduction
to mirror symmetry and its applications, we like to review in the
introduction
some general concepts of
string theory in view of mirror symmetry.
Throughout the text we have to assume some
familiarity with strings theory and conformal field theory
\ref\gswlt{For introductions, see e.g. M. Green, J. Schwarz and E.
Witten,
{\sl Superstring Theory}, Vols. I and II, Cambridge University Press,
1986;
D. L\"ust and S. Theisen, {\sl Lectures on String Theory},
Springer Lecture Notes in Physics, Vol. 346, 1990;
M. Kaku, {\sl Introduction to Superstrings},
Springer 1988}.

Some basic properties of closed string theory are best discussed
in the geometrical approach, i.e. by looking at the classical
$\sigma$-model action. It is defined by a map $\Phi$ from a compact
Riemann surface $\Sigma_g$ of genus $g$
(the world-sheet with metric $h_{\a\b}$) to the target space
$X$ (the space-time)
$\Phi:\Sigma_g \rightarrow X$ and an action $S(\Phi,G,B)$,
which may be viewed as the action of a two dimensional field theory.
The latter  depends on the dynamical field $\Phi$,
whereas the metric $G$ of $X$ and an antisymmetric
tensorfield $B$ on $X$ are treated as background fields.
As the simplest example one may take a bosonic action, which
reads
\eqn\lag{
S={1\over 2 \pi \alpha'}\int_{\Sigma_g} d^2\sigma\sqrt{h}\Bigl(
h^{\a\b}G_{ij}(\phi)\p_\a\phi^i\p_\b\phi^j+\e^{\a\b}
B_{ij}(\phi)\p_\a\phi^i\p_\b\phi^j\,+\,\dots\Bigr)
}
where $\phi^i$ ($i=1,\dots,{\rm dim}(X)$)
and $\sigma^\a$ ($\a=1,2$)
are local coordinates on $\Sigma_g$ and $X$
respectively. The dots indicate further terms, describing the
coupling to other background fields such as the
dilaton and gauge fields.
The first quantized string theory can then be perturbatively
defined in terms of a path integral
as\foot{We always assume that we quantize in the critical dimension.
Integration over the world-sheet metric can then be converted
to an integration over the moduli space ${\cal M}_g$ with suitable
measure which, for general correlation functions, depends on the
number of operator insertions
\ref\verlinde{E. Verlinde, H.Verlinde, {\sl Lectures
on String Perturbation Theory}, in {\sl Superstrings 88}, Trieste
Spring
School, Ed. Greene et al. World Scientific (1989) 189}.
By ${\cal S}(X)$ we mean the generating function of all
correlators of the string theory on $X$.}
\eqn\st{{\cal S}(X)=\sum_{g}\int_{{\cal M}_g} \int D\Phi e^{i
S(\Phi,G,B,\dots)}}

For a particular background to provide a classical string vacuum, the
sigma model based on it has to be conformally invariant
\ref\cfmp{C. Callan, D. Friedan, E. Martinec and M. Perry,
\npb262(1985)593, A. Sen, Phys. Rev. D32 (1985) 2102 and \prl55(1985)1846}.
This means that the energy-momentum tensor, including corrections from
$\sigma$-model
loops, must be traceless, or, equivalently, that the
$\beta$-functions
must vanish. Vanishing of the dilaton $\beta$-function demands that
we
quantize in the critical dimension, whereas the $\beta$-functions
associated to the metric and the anti-symmetric tensor impose
dynamical equations for the background, in particular that
is has (to lowest order in $\a'$)
to be Ricci flat, i.e. that the metric
satisfy the vacuum Einstein equations\foot{In fact, this point is
subtle,
as for $X$ a Calabi-Yau manifold (cf. section two),
higher ($\geq4$) $\sigma$-model loop effects modify the
equations of motion for the background. It can however be shown
\ref\ns{D. Nemeschanski and A. Sen, \plb178(1986)365}
that $\sigma$-models for Calabi-Yau compactifications are
conformally invariant and that by
means of a (non-local) redefinition of the metric one
can always obtain $R_{i\bar\jmath}=0$
as the equations of motion for the background.}.

Since we are dealing with strings, it is not the classical
geometry (or even topology) of $X$ which is relevant. In fact,
path integrals such as \st~ are related to the loop space of $X$.
Much of the attraction
of string theory relies on the hope that the modification of the
concept of classical geometry to string geometry at very small
scales will lead to interesting effects and eventually
to an understanding of physics in this range.
At scales large compared to the scale of the loops (which is related
to
$\a'$) a description in terms of point particles should be valid and
one should recover classical geometry.
The limit in which the classical description is valid is referred to
as the large radius limit.

One particular property of strings as compared to ordinary point
particles
is that there might be more than one manifold $X$ which leads
to identical theories; i.e. ${\cal S}(X_1)={\cal S}(X_2)$\foot{This
resembles the situation of quantized point
particles on so called isospectral manifolds. However in string theory the
invariance is more fundamental, as no experiment can be performed
to distinguish between $X_1$ and $X_2$.}.
The case where two manifolds have just
different {\it geometry} is usually referred to as duality symmetry
\ref\gpr{For review, see A. Giveon, M. Porrati and E. Rabinovici,
{\sl Target Space Duality in String Theory}, hep-th 9401139, to
be published in Phys. Rep.}.
%If they have different topology one is dealing with mirror symmetry.
Mirror symmetry
\ref\dixon{L. Dixon in {\sl Proc. of the 1987 ICTP Summer Workshop in
High Energy Physics and Cosmology}, Trieste, ed. G. Furlan et al.,
World Scientific}
\ref\lvw{W. Lerche, C. Vafa and N. Warner, \npb324(1989)427}
\ref\gp{B. Greene and M. Plesser, \npb338(1990)15}
\ref\mirror{{\sl Essays on Mirror Manifolds}
(ed. S.-T. Yau), Int. Press, Hong Kong, 1992}.
relates, in the generic case, identical string theories
on {\it topologically} different manifolds.
These symmetries are characteristic features of string geometry. For
the case of mirror symmetry, which is the central topic of these
lectures, this will become evident as we go along.

So far the analysis of (parts of) ${\cal S}(X)$ can be explicitly
performed only for very simple target spaces $X$, such as the torus
and orbifolds. Much of our understanding about the relation between
classical and string geometry is derived from these examples.
As a simple example we want to discuss compactification of the
bosonic string on a two dimensional torus
\ref\dvv{R. Dijkgraaf,
E. Verlinde and H. Verlinde, Utrecht preprint THU-87/30;
A. Shapere and F. Wilczek \npb320(1989)669; A. Giveon, E. Rabinovici,
G. Veneziano, \npb322(1989)169}; for review, see also
\ref\ferraratheisen{S. Ferrara and S. Theisen,
{\sl Moduli Spaces, Effective Actions and Duality Symmetry
in String Compactifications}, in Proc. of the Third Hellenic
School on Elementary Particle Physics, Corfu 1989,
Argyres et al.(eds), World Scientific 1990}.
This also allows us
to introduce some concepts which will appear in greater generality
later on.

The torus $T^2=\IR^2/\Gamma$ is defined by a two-dimensional lattice
$\Gamma$ which is generated by two basis vectors $e_1$ and $e_2$.
The metric, defined by $G_{ij}=e_i\cdot e_j$, has three independent
components and the antisymmetric tensor $B_{ij}=b\epsilon_{ij}$
has one component. For any values of the altogether four real
components
does one get a consistent string compactification. We thus have four
real moduli for strings compactified on $T^2$. We can combine them
into two complex moduli as follows:
$\sigma={|e_1|\over|e_2|}e^{i\phi}$ and $\tau=2(b+iA)$
where $\phi$ is the angle between the two basis vectors which we can,
without loss of generality, choose to be $0\leq\phi\leq\pi$ and
$A=\sqrt{|\det G|}>0$ is the area of the unit cell of $\Gamma$.
$\sigma$ parameterizes different complex structures on the torus
and is usually called the Teichm\"uller parameter. The imaginary part
of $\tau$ parameterizes the K\"ahler structure of the torus. We have used
the antisymmetric tensor field to complexify the K\"ahler modulus.
The role the two moduli play is easily recognized if one
considers $ds^2=G_{ij}dx^i dx^j\equiv\tau_2 dz d\bar z$
where the relation between
the real coordinates $x_i$ and the complex coordinates $z,\bar z$
only involves $\sigma$ but not $\tau$.

If one now considers the spectrum of the theory, one finds various
symmetries. They restrict the moduli space of the compactification
which is naively just two copies of the upper half complex plane.
With the definition of the left and right momenta
$$
p_L^2={1\over\sigma_2\tau_2}|(m_1-\sigma m_2)-\tau(n_2+\sigma
n_1)|^2\,, \quad
p_R^2= {1\over\sigma_2\tau_2}|(m_1-\sigma m_2)-\bar\tau(n_2+\sigma
n_1)|^2
$$
the spectrum of the energy and conformal
spin eigenvalues can be written as
\eqn\spectrum{
m^2=p_L^2+p_R^2\,+\,N_L+N_R-2\, ,\quad
s=p_L^2-p_R^2+N_L-N_R}
where $n_i$ and $m_i$ are winding and momentum quantum numbers,
respectively,
$N_{L,R}$ are integer oscillator contributions and the last term
in $m^2$ is from the zero point energy.
The symmetries of the theory are due to invariance of
\spectrum~under the transformation
$(\sigma,\tau)\mapsto(\tau,\sigma)$,
$(\sigma,\tau)\mapsto  (-\bar \sigma,-\bar \tau)$,
$(\sigma,\tau)\mapsto(\sigma+1,\tau)$  and
$(\sigma,\tau)\mapsto(-{1\over \sigma}, \tau)$ accompanied by
a relabeling and/or interchange of the winding and momentum quantum numbers.
The transformation which interchanges the two types of
moduli generates in fact what we call mirror symmetry.
The torus example is however too simple to exhibit a change of
topology as it is its own mirror.
The transformations which reflect the string property are those which
require an interchange of momentum and winding modes. The last three of the
generators given above are not of this type and they are also present
for the point particle moving on a torus (then $n_1=n_2=0$).
It is the addition of the mirror symmetry generator which introduces
the stringy behavior. Interchange of the two moduli must
be accompanied by $n_2\leftrightarrow m_2$. Composing
the mirror transformation with some of the other generators
given above, always involves interchanges of winding and momentum
quantum numbers.
E.g. for the transformation $\tau\to-{1\over\tau}$ we have to redefine
$m_1\to n_2,m_2\to-n_1,n_2\to-m_1,n_1\to m_2$ and if we set
$b=0$ then this transformation identifies compactification
on a torus of size $A$ to compactification on a torus of
size $1/A$.
Integer shifts of $\tau$ are discrete Peccei-Quinn symmetries.
One can show that the interactions (correlation functions)
are also invariant under these symmetries.

These lectures deal with mirror symmetry of strings
compactified on Calabi-Yau spaces. In section two we will review
some of the main features of Calabi-Yau compactifications, in
particular
the correspondence of complex structure and K\"ahler moduli with
elements of the cohomology groups $H^{2,1}$ and $H^{1,1}$,
respectively.
For $X$ and $X^*$ to be a mirror pair of Calabi-Yau manifolds
(we will use this notation throughout) one needs that
$h^{p,q}(X)=h^{3-p,q}(X^*)$ (for three dimensional
Calabi-Yau manifolds this is only non-trivial for $p,q=1,2$). The {\it
mirror
hypothesis} is however much more powerful since it states that the
string theory on $X$ and $X^*$ are identical,
i.e. ${\cal S}(X)={\cal S}(X^*)$. In particular it implies that one
type of
couplings on $X$ can be interpreted as another type of couplings
on $X^*$ after exchanging the role of the complex structure and the
K\"ahler moduli.

In section three we give a description of Calabi-Yau
compactification in terms of symmetric $(2,2)$ superconformal
field theory. The moduli of the Calabi-Yau space correspond
to exactly marginal deformations of the conformal field theory.
They come in two classes. Mirror symmetry  appears as a
trivial statement, namely as the change of relative sign, which is
pure convention, of two $U(1)$ charges \dixon\lvw.
By this change the two classes
of marginal perturbations get interchanged.
This does not change the
conformal field theory and thus leads to the same string vacuum.
In the geometrical interpretation this is however non-trivial,
as it entails the mirror hypothesis which implies the existence of
pairs of topologically different manifolds with identical string
propagation.

We will apply mirror symmetry to the computation of Yukawas couplings
of charged matter fields. They come in two types, one easy to compute
and the other hard to compute. On the mirror manifold, these two
couplings change role. What one then does is to compute the easy ones
on either manifold and then map them to one and the same manifold via the
so-called mirror map. In this way one obtains both types of couplings.
This will be explained in detail in section
six. Before getting there we will show how to construct mirror pairs
and how to compute the easy Yukawa couplings. This will be done in sections
four and five. In section seven we present an example in detail, where
the concepts introduced before will be applied. In the final section
we draw some conclusions.

Before continuing to section two, let us give a brief guide to the
literature. The first application of mirror symmetry was given in
the paper by Candelas, de la Ossa, Green and Parkes
\ref\cdgp{P. Candelas, X. De la Ossa, P. Green
and L. Parkes, \npb359(1991)21}
where the simplest Calabi-Yau manifold
was treated, the quintic in $\IP^4$ which has only one K\"ahler
modulus.
Other one-moduli examples were covered in
\ref\morrison{D.Morrison, {\sl Picard-Fuchs Equations and
Mirror Maps for Hypersurfaces}, in {\sl Essays on Mirror Manifolds}
(ed. S.-T. Yau), Int. Press, Hong Kong, 1992;
A. Klemm and S. Theisen, \npb389(1993)153;
A. Font, \npb391(1993)358}
(for hypersurfaces) and in
\ref\lt{A. Libgober and J. Teitelbaum, {\sl Duke Math. Journ., Int.
Res.
Notices} 1 (1993) 29; A. Klemm and S. Theisen, {\sl Mirror Maps and
Instanton Sums for Complete Intersections in Weighted Projective
Space},
preprint LMU-TPW 93-08}
(for complete intersections).
Models with several moduli were examined in refs.
\ref\cdfkm{P. Candelas, X. de la Ossa, A. Font, S. Katz and
D. Morrison, {\sl Mirror Symmetry for Two Parameter Models I},
preprint CERN-TH.6884/93}
\ref\hktyI{S. Hosono, A. Klemm, S. Theisen and S.-T. Yau,
{\sl Mirror Symmetry, Mirror Map and Applications to
Calabi-Yau Hypersurfaces}, to appear in Comm. Math. Phys.}
(two and three moduli models) and
\ref\hktyII{S. Hosono, A. Klemm, S. Theisen and S.-T. Yau,
{\sl Mirror Symmetry, Mirror Map and Applications to Complete
Intersection Calabi-Yau Spaces}, preprint LMU-TPW-94-03
(to appear)}.
Other references, especially to the mathematical literature, will
be given as we go along.
A collection of papers devoted to various aspects of mirror symmetry
is \mirror.
Some of the topics and results
to be discussed here are also contained in \cdfkm
\ref\agm{P. Aspinwall, B. Greene and D. Morrison, {\sl Space-Time
Topology Change: the Physics of Calabi-Yau Moduli Space},
preprint IASSNS-HEP-93-81; {\sl Calabi-Yau Moduli Space, Mirror
Manifolds
and Space-Time Topology Change in String Theory}, preprint
IASSNS-HEP-93-38;
{\sl The Monomial-Divisor Mirror Map}, preprint IASSNS-HEP-93-43}
\ref\bcofhjq{P. Berglund, P. Candelas, X. de la Ossa, A. Font, T.
H\"ubsch,
D. Jancic and F. Quevedo, {\sl Periods for Calabi-Yau and
Landau-Ginsburg
Vacua}, preprint CERN-TH.6865/93;
P. Berglund, E. Derrich, T. H\"ubsch and D. Jancic,
{\sl On Periods for String Compactification}, preprint HUPAPP-93-6}
\ref\per{P. Berglund and S. Katz, {\sl Mirror Symmetry for
Hypersurfaces in
Weighted Projective Space and Topological Couplings}, preprint
IASSNS-HEP-93-65}.
These notes draw however most heavily from our own papers on the
subject.

%%%%%%%%%%%%%%%%%%%%%%%%%%%%%%%%%%%%%%%%%%%%%%%%%%%%%%%%%%%%%%%%%%%%
\newsec{Strings on Calabi-Yau Manifolds}
%%%%%%%%%%%%%%%%%%%%%%%%%%%%%%%%%%%%%%%%%%%%%%%%%%%%%%%%%%%%%%%%%%%%
One of the basic facts of string theory is the existence of a
critical dimension, which for the heterotic string, is ten.
To reconciliate this with the observed four-dimensionality
of space-time, one makes the compactification Ansatz
that the ten-dimensional
space-time through which the string moves has the direct product form
$X_{10}=X_{4}\times X_6$ where $X_6$ is a six-dimensional compact
internal manifold, which is supposed to be small, and $X_4$
is four-dimensional Minkowski space. If one imposes the
`phenomenologically' motivated condition that the theory has
$N=1$ supersymmetry in the four uncompactified dimensions,
it was shown in
\ref\chsw{P. Candelas, G. Horowitz, A. Strominger and E. Witten,
\npb258(1985)46}
that $X_6$ has to be a so-called Calabi-Yau manifold
\ref\cy{Easily accessible introductions to Calabi-Yau manifolds
are G. Horowitz, {\sl What is a Calabi-Yau Space?} in {\sl Unified
String Theories}, M. Green and D. Gross, editors, World Scientific
1986
and P. Candelas, {\sl Introduction to Complex Manifolds}, Lectures at the
1987 Trieste Spring School, published in the proceedings.
For a rigorous mathematical treatment we refer to
A.L. Bessis, {\sl Einstein Manifolds}, Springer 1987;
P. Griffiths and J. Harris, {\sl Principles of Algebraic Geometry},
John Wiley \& Sons 1978;
K. Kodaira, {\sl Complex Manifolds and Deformation of Complex Structures},
Springer 1986}
\ref\huebsch{A good and detailed introduction to Calabi-Yau manifolds
is the book by
T. H\"ubsch, {\sl Calabi-Yau Manifolds}, World Scientific 1991}.

{\it Def.\foot{Here and below we
restrict ourselves to the three complex-dimensional case.}:
A Calabi-Yau manifold is a compact K\"ahler manifold with
trivial first Chern class.}

The condition of trivial first Chern class on a compact
K\"ahler manifold is, by Yau's theorem,
equivalent to the statement that they admit a
Ricci flat K\"ahler metric. The necessity is easy to see, since
the first Chern class $c_1(X)$ is represented by the 2-form
${1\over 2\pi}\rho$ where $\rho$ is the Ricci two form, which is the
2-form associated to the Ricci tensor of the K\"ahler metric:
$\rho=R_{i\bar\jmath}dz^i\wedge d\bar z^{\bar\jmath}$.
Locally, it is given by
$\rho=-i\p\bar\p\log\det ((g_{i\bar\jmath}))$.
One of the basic properties of Chern classes is their
independence of the choice of K\"ahler metric; i.e.
$\rho(g')=\rho(g)+d\a$.
If now $\rho(g)=0$, $c_1(X)$ has to be trivial. That this is also
sufficient was conjectured by Calabi and proved by
Yau
\ref\yau{S.T. Yau, Proc. Natl. Acad. Sci. USA 74 (1977), 1798;
Comm. Pure Appl. Math.  31 (1978) 339}.

Ricci flatness also implies that the holonomy group is
contained in $SU(3)$ (rather than $U(3)$; the $U(1)$ part is
generated
by the Ricci tensor $R_{i\bar\jmath}
=R_{i\bar\jmath k}{}^k$). If the holonomy is $SU(3)$
one has precisely $N=1$ space-time supersymmetry. This is what we
will
assume in the following. (This condition e.g. excludes the
six-dimensional torus $T^6$, or $K_3\times T^2$, which would lead
to extended space-time supersymmetries.)

Another consequence of the CY conditions is the existence
of a unique nowhere vanishing covariantly constant
holomorphic three form, which
we will denote by $\Omega=\Omega_{ijk}d z^i\wedge dz^j
\wedge dz^k\,\,(i,j,k=1,2,3)$, where $z^i$ are local complex
coordinates
of the CY space. Since $\Omega$ is a section of the canonical
bundle\foot{The canonical bundle is the highest (degree ${\rm
dim}_{\IC} (X)$)
exterior power of the holomorphic cotangent bundle $T_X^*$.},
vanishing of the first Chern class is equivalent to the
triviality of the canonical bundle.

A choice of complex coordinates defines a complex structure. The transition
functions on overlaps of coordinate patches are holomorphic functions.
There are in general families of possible complex structures on
a given CY manifold. They are parameterized by the so-called
complex structure moduli.
Using Kodaira-Spencer defomation theory
\ref\kns{K. Kodaira, L. Nirenberg and D. C. Spencer,
\aom68(1958)450},
it was shown in
\ref\tian{G. Tian, in {\sl Mathematical Aspects of String
Theory}, ed. S. T. Yau, World Scientific, Singapore (1987)}
that for Calabi-Yau manifolds this parameter space is locally
isomorphic to
an open set in $H^1(X,T_X)$.
For algebraic varieties the deformation along elements of
$H^1(X,T_X)$ can often be described by deformations of the defining
polynomials (cf. section four).

In addition to the complex structure moduli there are also the
K\"ahler moduli. They parameterize the possible K\"ahler forms.
A K\"ahler form is a real closed
$(1,1)$ form $J=\omega_{i\bar\jmath}dz^i\wedge d\bar z^{\bar\jmath}$
(with the associated
K\"ahler metric $g_{i\bar\jmath}=i\omega_{i\bar\jmath}$)
which satisfies the positivity conditions
\eqn\kaehlerconditions{\int_{C}J>0\,,\quad\int_S J^2>0\,,\quad
                       \int_X J^3>0}
for all curves $C$ and surfaces $S$ on the CY manifold $X$.
Since ${1\over 3!}J^3$ is the volume form on $X$, one concludes
that the K\"ahler form cannot be exact and consequently
one has ${\rm dim}(H^{1,1}(X))\equiv h^{1,1}\geq1$ for
$X$ K\"ahler.
If there are more than one harmonic $(1,1)$ forms on $X$, i.e.
if $h^{1,1}>1$, then
$\sum_{a=1}^{h^{1,1}}t'_a h_a$, $t'\in\IR$, with
$h_a\in H^{1,1}_{\bar\p}(X)$ will define a K\"ahler class,
provided the K\"ahler moduli lie within the
so-called K\"ahler cone, i.e. \kaehlerconditions~ is satisfied.

{}From the local expression of the Ricci form it follows that it
depends on the complex structure and on the volume form of the
K\"ahler metric. The question now arises whether by changing
the K\"ahler form and the complex structure Ricci flatness is
preserved.
This means that the moduli of the CY manifold
must be associated with deformations of the
{\it Ricci flat} K\"ahler metric: $\delta g_{i\bar\jmath}$ with
K\"ahler deformations and $\delta g_{ij}$ and
$\delta g_{\bar\imath\bar\jmath}$ with deformations of the
complex structure. If one examines the condition
$\rho(g+\delta g)=0$ one finds that $\delta g_{i\bar\jmath}dz^i
\wedge d\bar z^{\bar\jmath}$
is harmonic, i.e. we can expand it as
$\delta g_{i\bar\jmath}dz^i\wedge d\bar z^{\bar\jmath}
=\sum_{a=1}^{h^{1,1}} \delta t'_a h_a$.
Likewise, $\Omega_{ij}{}^{\bar l}\delta g_{\bar l\bar k}dz^i\wedge
dz^j\wedge d\bar z^{\bar k}=
\sum_{\a=1}^{h^{2,1}}\delta\lambda_\a b_\a$, with
$b_a\in H^{2,1}(X)$.
Here we have employed the unique $\Omega$.

%One may show that for a Calabi-Yau manifold the
%Hodge diamond is
%\eqn\hodgediamond{
%\atrix{h^{0,0}\cr
%        h^{1,0}\qquad h^{0,1}\cr
%        h^{2,0}\qquad h^{1,1}\qquad h^{0,2}\cr
%        h^{3,0}\qquad h^{2,1}\qquad h^{1,2}\qquad h^{0,3}\cr
%        h^{3,1}\qquad h^{2,2}\qquad h^{1,3}\cr
%        h^{3,2}\qquad h^{2,3}\cr
%        h^{3,3}\cr}\quad\equiv\quad
%\matrix{1\cr
%        0\qquad 0\cr
%        0\qquad h^{1,1}\qquad 0\cr
%        1\qquad h^{2,1}\qquad h^{1,2}\qquad 1\cr
%        0\qquad h^{2,2}\qquad 0\cr
%        0\qquad 0\cr
%        1\cr}}
One can show that the only independent non trivial Hodge numbers of
CY
manifolds are $h^{0,0}=h^{3,0}=1$ and $h^{1,1}$ and $h^{2,1}$
depending on the particular manifold. In addition we have
$h^{p,q}=h^{q,p}$ (complex conjugation), $h^{p,q}=h^{3-p,3-q}$
(Poincar\'e duality) and $h^{0,p}=h^{0,3-p}$ (isomorphism via
$\Omega$).
For K\"ahler manifolds the Betti numbers are
${b_r=\sum_{p+q=r}h^{p,q}}$ and the Euler number is thus
$\chi(X)=\sum_{r=0}^{{\rm dim}(X)} b_r$.

If we were geometers we would
only be interested in the deformations of the {\it metric} and
the number of (real) moduli would be
$h^{1,1}+2 h^{2,1}$. However, in  string theory compactified on CY
manifolds we have additional massless scalar degrees of freedom from
the
non-gauge sector, namely those coming from the (internal components
of the) antisymmetric tensor field $B_{i\bar\jmath}$. As a result of
the
equations of motion it is a harmonic $(1,1)$ form and its changes
can thus be parameterized as
$\delta B_{i\bar\jmath}dz^i\wedge d\bar z^{\bar\jmath}
=\sum_{a=1}^{h^{1,1}}\delta t''_a h_a$ where
$t''_a$ are real parameters. One combines
$(i\delta g_{i\bar\jmath}+
\delta B_{i\bar\jmath})dz^i\wedge dz^{\bar\jmath}
=\sum_{a=1}^{h^{1,1}} \delta t_a h_a$ where now the $t_a=t_a''+it_a'$
are complex parameters.
This is referred to as the complexification of the K\"ahler cone.

Recall (and see below)
that for strings on CY manifolds, the massless sector
of the theory consists of a universal sector, containing the
graviton,
an antisymmetric tensor field (by duality related to the axion) and
a dilaton, and a matter sector with $h^{1,1}$ $27$-plets and
$h^{2,1}$ $\overline{27}$-plets of $E_6$ and a certain number of
$E_6$ singlets. $E_6$ invariance restricts the possible Yukawa
couplings
to the following four types: $\bra 27^3\ket$,
$\bra\overline{27}^3\ket$,
$\bra 27\cdot\overline{27}\cdot 1\ket$ and $\bra 1^3\ket$. In the
following
we will only treat the former two couplings
\ref\sw{A. Strominger, \prl55(1985)2547, and in {\sl Unified String
Theory},
eds. M. Green and D. Gross, World Scientific 1986;
A. Strominger and E. Witten, \cmp101(1985)341}.
Not much is known about the remaining two\foot{except for cases
in which the corresponding conformal field theory can be treated exactly,
e.g. for Calabi-Yau spaces with an toroidal orbifold limit
they can be calculated for the untwisted sector at the orbifold point
\ref\dfms{L. Dixon, D. Friedan, E. Martinec and S, Shenker,
          \npb282(1987)13;
           S. Hamidi and C. Vafa,\npb279(1987)465;
           J. Lauer, J. Mas and H. P. Nilles  \npb351 (1991) 353;
           S. Stieberger, D. Jungnickel, J. Lauer and
           M. Spalinski, Mod. Phys. Lett. A7 (1992) 3859;
           J. Erler, D. Jungnickel, M. Spalinski and
           S. Stieberger, \npb397(1993)379}
and for Calabi-Yau spaces with Gepner
\ref\gepner{D. Gepner, \plb199(1987)380 and \npb311(1988)191}
model interpretion at one point (the Gepner point) in moduli space.}.

One considers the coupling
of two fermionic and one bosonic field (Yukawa coupling)
in the ten-dimensional field theory. All these fields are
in the fundamental (248) representation of $E_8$, the gauge
group of the uncompactified heterotic string\foot{We do not
consider the second $E_8$ factor here. It belongs to the
so-called hidden sector.}.
(The bosonic field is the $E_8$ gauge field.) One then
expands the fields in harmonics
on the internal CY manifold and arrives at couplings which factorize
into two terms: one is a cubic coupling of three fields on the
four-dimensional Minkowski space and the other an overlap integral
over the internal manifold of three zero-modes
(we are interested in massless fields)
of the Dirac and the wave operators, respectively.
The second factor is the effective Yukawa coupling of the
four-dimensional field theory. Under $E_8\supset E_6\times SU(3)$
we have the decomposition
$248=(27,3)\oplus(\overline{27},\bar 3)
\oplus(1,8)\oplus(78,1)$. The $(78,1)$ gives the $E_6$ gauge fields
and the $(1,8)$ is the spin connection which has been identified with the
$SU(3)$ part of the $E_8$ gauge connection \chsw.

The four-dimensional matter fields transform as $27$ and
$\overline{27}$
of $E_6$ and the zero modes in the internal CY manifold carry a
$SU(3)$ index in the $3$ and $\bar3$ representations, respectively.
Group theory then tells us that there are
two different kinds of Yukawa couplings among the charged matter
fields:
$\bra 27^3\ket$ and $\bra\overline{27}^3\ket$.
The zero modes on the internal manifold
can be related to cohomology elements of the CY space and one finds
\sw~
that the two types of Yukawa couplings are of the form:
\eqn\yukawas{
\eqalign{\kappa_{abc}^{0\sst (27)}(X)\equiv \kappa_{abc}^0(X)
&=\int_X h_a\wedge h_b\wedge h_c,\qquad\qquad\qquad\,
a,b,c=1,\ldots,h^{1,1}\cr
\kappa_{\a\b\g}^{\sst{(\overline{27})}}(X)\equiv\bar\kappa_{\a\b\g}(X)
&
         =\int_X \Omega\wedge b_\a^i\wedge b_\b^j\wedge
b_\g^k\,\Omega_{ijk},
\qquad \a,\b,\g=1,\ldots,h^{2,1}}
}
where $h_a$ are the harmonic $(1,1)$ forms and
$b^i_\a=(b_\a)^i_{\bar\jmath}d\bar z^{\bar\jmath}$ are elements of
$H^1(X,T_X)$
which are related to the harmonic $(2,1)$ forms
via the unique element of $H^3(X)$:
$(b_\a)^i_{\bar\jmath}={1\over 2|\!|\Omega|\!|^2}
\Omega^{ikl}(b_\a)_{kl\bar\jmath}$,
$|\!|\Omega|\!|^2={1\over3!}\bar\Omega^{ijk}\Omega_{ijk}$.
Note that while the former couplings are purely topological the
latter
do depend on the complex structure (through $\Omega$).
Both types of cubic couplings are totally symmetric.
Note also that by the discussion above there is a one-to-one
correspondence
between charged matter fields and moduli: $27\leftrightarrow(1,1)$
moduli
and $\overline{27}\leftrightarrow(2,1)$ moduli\foot{This identification
is a matter of convention. Here we have identified the $\bar 3$ of
$SU(3)$ with a holomorphic tangent vector index. The $3$ of $SU(3)$
is a holomorphic cotangent vector index and one uses
(Dolbeault theorem) $H^1(X,T_X^*)\simeq H^{1,1}(X)$.}.

These results for the couplings have been derived in the
(classical) field theory
limit and do not yet incorporate the extended nature of strings.
This issue will be taken up next.

In general, to compute the string Yukawa couplings, one has to
take into account sigma model and string perturbative and
non-perturbative effects.
One can show that both types of Yukawa couplings
do not receive corrections from sigma model loops and string loops.
The couplings $\bar\kappa$ are, in fact, also unmodified by
non-perturbative effects on the world-sheet, which, due to absence
of string loop corrections, is just the sphere
\ref\digr{J. Distler and B. Greene, \npb309(1988)295}.
The couplings $\kappa^0$ do
however receive corrections from world-sheet instantons
\ref\dsww{M. Dine, N. Seiberg, X-G. Wen and E. Witten,
\npb278(1987)769,\npb289(1987)319}.
These are non-trivial holomorphic embeddings of the world-sheet
$\Sigma_0\simeq\IP^1$
in the CY manifold. In algebraic geometry they are known as rational
curves $C$ on $X$. Then there are still non-pertubative string-effects,
i.e. possible contributions from infinite genus world-sheets. We do not
know anything about them and will ignore them here. The possibility to
incorporate them in the low-energy effective action has been discussed in
\ref\flst{S. Ferrara, D. L\"ust, A. Shapere and S. Theisen,
\plb225(1989)363}.

The couplings in eq.\yukawas~ are computable in classical algebraic
geometry, and, were they the whole truth to the Yukawa
couplings of strings on CY manifolds, they would blatantly contradict
the mirror hypothesis: $\kappa^0$ is independent of moduli whereas
$\bar\kappa$ depends on the complex structure moduli. In fact, the mirror
hypothesis states that the full $\bra 27^3\ket$ couplings on the manifold $X$
depend on the K\"ahler moduli in such a way that they are related to the
$\bra\overline{27}^3\ket$ couplings on the mirror manifold $X^*$ via the
mirror map. The main topic of these lectures is to explain what this means
and to provide the tools to carry it through.
The dependence on the K\"ahler moduli is the
manifestation of string geometry and is solely due to the extended
nature of strings.

When computing the Yukawa-couplings in conformal field theory as
correlation
functions of the appropriate vertex operators,
inclusion of the non-perturbative $\sigma$-model effects means that
in the
path-integral one has to sum over all holomorphic embeddings of the
sphere in $X$. This is in general not feasible since it requires
complete
knowledge of all possible instantons and their moduli spaces. In
fact, this
is where mirror symmetry comes to help.

We have thus seen that, modulo the remark on non-perturbative string
effects,
the Yukawa couplings are
$\bar\kappa(X)$ and for the instanton corrected couplings
$\kappa(X)$ one expects an expansion of the form
($\imath: C\hookrightarrow X$)
\eqn\instantonsum{
\kappa_{abc}=\kappa^0_{abc}+\sum_C\int_C \imath^*(h_a)
\int_C \imath^*(h_b)\int_C \imath^*(h_c)
{e^{2\pi i\int_C\imath^*(J(X))}\over 1-e^{2\pi
i\int_C\imath^*(J(X))}},
}
which generalizes the Ansatz made in \cdgp,
which led to a successfull prediction for the numbers of instantons
on the quintic hypersurface in $\IP^4$, to the multi-moduli case.
This ansatz was justified in  ref.
\ref\am{P. Aspinwall and D. Morrison, \cmp151(1993)245} in the
framework of topological sigma models
\ref\topsig{E. Witten, \cmp118(1988)411}.
The sum is over all instantons $C$ of the sigma model based on $X$
and the denominator takes care of their multiple covers.
$J$ is the K\"ahler form on $X$.

One sees from \instantonsum~ that as we go `far out' in the K\"ahler
cone
to the `large radius limit', the instanton corrections get
exponentially
supressed and one recovers the classical result.

%%%%%%%%%%%%%%%%%%%%%%%%%%%%%%%%%%%%%%%%%%%%%%%%%%%%%%%%%%%%%%%%%%%%
\newsec{Superconformal Field Theory and CY Compactification}
%%%%%%%%%%%%%%%%%%%%%%%%%%%%%%%%%%%%%%%%%%%%%%%%%%%%%%%%%%%%%%%%%%%%

Let us now turn to an alternative view of string compactification. We
recall that the existence of a critical dimension is due to the
requirement that the total central charge
(matter plus ghost) of the Virasoro algebra
vanishes. The critical dimensions for the bosonic and fermionic
strings then follow from the central charges of the Virasoro
algebras generated by the reparametrization and local $n=1$
world-sheet
supersymmetry ghost systems, which are $c=-26$ and
$\hat c={2\over 3}c=-10$,
respectively. If we want a four-dimensional Minkowski space-time,
we need four left-moving free bosonic fields and four right-moving
free chiral superfields, contributing $(\bar c,c)=(4,6)$ to the
central
charge. (Barred quantities refer to the left-moving sector.)
Compactification might then be considered as an internal
conformal field theory with central charge $(\bar c_{\rm int},c_{\rm
int})
=(22,9)$. There are however internal conformal field theories which
satisfy all consistency requirements (e.g. modular invariance,
absence of dilaton tadpoles, etc.)
which do not allow for a geometric interpretation as
compactification. In the
case of CY compactification, the internal conformal field theory
is of a special type. The left moving central charge splits into
a sum of two contributions, $\bar c_{\rm int}=22=13+9$, where the
first part is due to a $E_8\times SO(10)$ gauge sector
(at level one; $E_8\times SO(10)$ is
simply laced of rank 13). The remaining contribution combines with
the right-moving part to a symmetric (i.e. the same for both left and
right movers) $(\bar n,n)=(2,2)$ superconformal field theory
with central charges $(9,9)$.
A right moving (global) $n=2$ extended algebra is necessary
for space-time supersymmetry
\ref\bdfm{T. Banks, L. Dixon, D. Friedan and E. Martinec,
\npb284(1988)613},
whereas the symmetry between left and
right movers are additional inputs which
allow for the geometrical interpretation as CY compactification with
the spin connection embedded in the gauge connection \chsw.

The fact that we have a left as well as a right moving extended
superconformal symmetry will be crucial for mirror symmetry.
Before explaining this, let us briefly mention the relevant features
of the $n=2$ superconformal algebra
\ref\ademolloetal{M. Ademollo et al., \npb11(1976)77 and
\npb114(1976)297
and \plb62(1976)105}.
It has four generators, the bosonic spin two energy momentum tensor $T$,
two fermionic spin $3/2$ super-currents
$T_F^\pm$ and a bosonic spin one current $J$
which generates a $U(1)$ Kac-Moody algebra.
If we expand the fields in modes as $T(z)=\sum L_n z^{-n-2}$,
$T_F^\pm(z)=\sum G_r^\pm z^{-r-3/2}$  and $J(z)=\sum J_n z^{-n-1}$
the algebra takes the form ($G_r^+=(G_{-r}^-)^\dagger$)
\eqn\scfalgebra{
\eqalign{\(L_n,L_m\)=(n-m)&L_{n+m}+{c\over
12}(n^3-n)\delta_{n+m,0}\cr
         \{G_r^\pm,G_s^\mp\}=2L_{r+s}\pm(r-s)J_{r+s}
         &+{c\over3}(r^2-{1\over 4})\delta_{r+s,0}\,,
         \quad \{G_r^\pm, G_s^\pm\}=0\cr
         \(L_n,G_r^\pm\)=({n\over2}-r)&G_{n+r}^\pm\,,\quad
         \(L_n, J_m\)=-mJ_{n+m}\cr
         \(J_n,J_m\)={c\over3}n&\delta_{m+n,0}\,,\quad
         \(J_n,G_r^\pm\)=\pm G^\pm_{n+r}}}
The moding of the fermionic generators is $r\in\Z$ in the Ramond ($R$)
and $r\in\Z+{1\over2}$ in the Neveu-Schwarz ($NS$)
sector. The finite dimensional
subalgebra in the $NS$ sector, generated by $L_{0,\pm1},\,J_0$ and
$G^\pm_{\pm1/2}$ is $OSp(2|2)$.
In a unitary theory we need
\eqn\unitary{
\bra\phi|\{G_r^\pm,G_{-r}^\mp\}|\phi\ket=2h\pm2rq+{c\over3}(r^2-{1\over4})\geq0
}
for any state with $U(1)$ charge $q$ (i.e. $J_0|\phi\ket=q|\phi\ket$)
and conformal weight $h$ (i.e. $L_0|\phi\ket=h|\phi\ket$).
Setting $r=0$ we thus find that in the $R$ sector $h\geq{c\over24}$
and
in the $NS$ sector (setting $r={1\over2}$) $h\geq{|q|\over2}$.

There is a one-parameter isomorphism of the algebra, generated by
$\U$, called spectral flow
\ref\schsei{A. Schwimmer and N. Seiberg, \plb184(1987)191}
\eqn\spectralflow{
\eqalign{\U L_n\Ui&=L_n+\theta J_n+{c\over 6}\theta^2\delta_{n,0}
         \quad\to\Delta h={3\over2c}\Delta(q^2)\cr
         \U J_n\Ui&=J_n+{c\over3}\theta\delta_{n,0}
         \quad\to\Delta q=+{c\over3}\theta\cr
         \U G^\pm_r\Ui&=G^\pm_{r\pm\theta}}}
States transform as $|\phi\ket\to \U|\phi\ket$.
For $\theta\in\Z+{1\over2}$ the spectral flow interpolates between
the
$R$ and the $NS$ sectors and for $\theta\in\Z$ it acts diagonally
on the two sectors.

We have two commuting copies of the $n=2$ algebra. The left moving
$U(1)$ current combines with the $SO(10)$ Kac-Moody algebra to
form an $E_6$ algebra at level one. Hence the gauge group
$E_6$ for CY compactifications. (The $E_8$ factor, which is also
present, will play no role here.)

Let us first consider the $R$ sector\foot{The following paragraphs draw
heavily from ref.\lvw.}.
Ramond ground states $|i\ket_R$ satisfy $G_0^\pm|i\ket_R=0$, i.e.
$\{G_0^+,G_0^-\}|i\ket_R=0$. From \unitary~ it follows
that $R$ ground states have conformal weight $h={c\over 24}$.
Under spectral flow by $\theta=\mp{1\over 2}$, the $R$ ground states
flow into {\it chiral/anti-chiral} primary states of the $NS$ sector.
They are primary states that satisfy the additional constraint
$G_{-1/2}^\pm|i\ket_{NS}=0$. (Recall that primary states are
annihilated by all positive modes of all generators of the algebra.)
It follows from \spectralflow~
that chiral/anti-chiral primary states satisfy $h=\pm{1\over 2}q$.
The $OSp(2|2)$ invariant $NS$ vacuum $|0\ket$
is obviously chiral and anti-chiral primary. Under spectral flow by
$\theta=\pm1$ it flows to a unique chiral (anti-chiral) primary field
$|\rho\ket\,(|\bar\rho\ket)$ with $h={c\over6}$ and $q=\pm{c\over3}$.
It follows from \unitary~
that for chiral primary fields $h\leq{c\over6}$.

We now look at the operator product of two chiral primary fields
$\phi_i(z)\phi_j(w)\sim\sum_k(z-w)^{h_k-h_i-h_j}\psi_k(w)$
where the $\psi_k$ are necessarily chiral but not necessarily
primary. $U(1)$
charge conservation requires $q_k=q_i+q_j$ and due to the inequality
$h\geq{|q|\over2}$ with equality for primary fields we conclude
that in the limit $z\to w$ only the chiral primaries survive.
They can thus be multiplied pointwise
and therefore form a ring ${\cal R}$ under multiplication:
$\phi_i\phi_j=
\sum_k c_{ij}{}^k\phi_k$ where the structure constants are functions
of the moduli (cf. below). This ring is called chiral
primary ring. The same obviously holds for anti-chiral fields forming
the anti-chiral ring.

Note that spectral flow by $\theta$ is merely a shift of the $U(1)$ charge
by ${c\over3}\theta$ and the accompanying change in the conformal weight.
Indeed, in terms of a canonically normalized boson
($\phi(z)\phi(w)=-\ln(z-w)+\dots$) we can express the $U(1)$ current as
$J(z)=\sqrt{c\over3}\p\phi(z)$. Any field\foot{Recall the correspondence
between states and fields: $|\phi\ket=\lim_{z\to0}\phi(z)|0\ket$.}
with $U(1)$ charge $q$ can be written as
$\phi_q=e^{i\sqrt{3\over c}q\phi}{\cal O}$ where ${\cal O}$ is neutral
under $U(1)$. The conformal weight of $\phi_q$ is
${3\over2c}q^2+h_{\cal O}$
and of $\phi_{q\!+\!{c\over3}\theta}$ it is
${3\over2c}(q\!+\!{c\over3}\theta)^2\!+\!h_{\cal O}$, in agreement with
\spectralflow. We thus find that the spectral flow operator
${\cal U}_\theta$ can
be written as ${\cal U}_\theta=e^{i\theta\sqrt{c\over3}\phi}$ and also
$\rho(z)={\cal U}_1(z)=e^{i\sqrt{c\over3}\phi(z)}$.

The foregoing discussion of course
applies separately to the left and right moving sectors and we in fact
have four rings: $(c,c)$ and $(a,c)$ and their conjugates $(a,a)$ and $(c,a)$.
Here $c$ stands for chiral and $a$ for anti-chiral.

We will now make the connection to our discussion of CY
compactification and set $(\bar c,c)=(9,9)$.
We have already mentioned that $N=1$ space-time supersymmetry for
the heterotic string requires $n=2$ superconformal symmetry for the
right movers. This is however not sufficient. The additional requirement
is that for states in the right-moving
$NS$ sector $q_R\in\ZZ$. The reason for this
is the following \bdfm. The operator ${\cal U}_{1\over2}(z)$ takes states
in the right-moving $NS$ sector to states in the right-moving $R$ sector
i.e. it transforms space-time bosons into space-time fermions
(and vice versa). In fact,
${\cal U}_{1\over2}(z)=e^{i{\sqrt3\over2}\phi(z)}$ is the internal part of the
gravitino vertex operator, which, when completed with its space-time and
super-conformal ghost parts, must be local with respect to all the fields
in the theory. When considering space-time bosons this
leads to the requirement
of integer $U(1)$ charges\foot{To see this, take the operator product
of the gravitino vertex operator (e.g. in the $1/2$ ghost picture)
$\psi_\a(z)=e^{-\varphi/2}S^\a e^{i{\sqrt3\over2}\phi}(z)$ $(S^\a$ is a
$SO(4)$ spin field) with a space-time boson (in the zero ghost picture)
with vertex operator $V=({\rm s.t.})e^{i{q\over\sqrt3}\phi}$. The
operator products of the spin field with the space-time parts
(s.t) are either local, in which case we need $q=2\ZZ+1$, or
have square root singularities, and we need $q=2\ZZ$.}
(which have to be in the range $-3,\dots,+3$)\foot{Space-time
supersymmetry and the existence of the unique state $|\rho\ket$
with $q={c\over3}$ thus requires that $c$ be an integer multiple of 3.}.
Since we are dealing with {\it symmetric} $(2,2)$ superconformal theories
both $q_L$ and $q_R$ must be integer for states in the $(NS,NS)$ sector.

We now turn to the discussion of the moduli of the Calabi-Yau
compactification. In an effective low energy field theory they are
neutral (under the gauge group $E_6\times E_8$) massless scalar fields
with vanishing potential (perturbative and non-perturbative)
whose vacuum expectation values determine
the `size' (K\"ahler moduli) and `shape' (complex structure moduli)
of the internal manifold. In the conformal field theory context they
parameterize the perturbations of a given conformal field theory by
exactly marginal operators.  Exactly marginal operators
can be added to the action without
destroying (2,2) super-conformal invariance of the theory.
Their multi-point correlation functions all vanish. In fact, one can show
\dsww\dixon
\ref\dineseiberg{M. Dine and N. Seiberg, \npb301(1988)357}
that there is a one-to-one correspondence between moduli of the (2,2)
superconformal field theory and
chiral primary fields with conformal weight
$(\bar h,h)=({1\over2},{1\over2})$.
The chiral primary fields are (left and right) (anti)chiral superfields
whose upper components (they survive the integral over chiral superspace)
have conformal weight (1,1) and are thus marginal.
The lower components with weights
$({1\over2},{1\over2})$ provide the internal part of the
charged matter fields ($27$ and $\overline{27}$ of $E_6$).
The gauge part on the left moving side and the space-time and
superconformal ghost parts on the right-moving side account
for the remaining half units of conformal weight for the massless
matter fields. We have thus a one-to-one correspondence between
charged matter fields and moduli: extended world-sheet supersymmetry
relates the $27$'s of $E_6$ to the marginal operators in the
$(c,c)$ ring and the $\overline{27}$'s of $E_6$ to the marginal
operators in the $(a,c)$ ring\foot{Note that this identification
is again a matter of convention. The arbitrariness here is due to a trivial
symmetry of  the theory under the flip of the relative sign of the
left and right $U(1)$ charges (cf. below).}.

We have encountered the four chiral rings. The fields in the four rings
can all be obtained
from the $(R,R)$ ground states via spectral flow. The additive structure
of the rings is therefore isomorphic, not however their multiplicative
structures.
They are in general very different.
The $(c,c)$ ring contains fields with $U(1)$ charges
$(q_L,q_R)=(+1,+1)$
whereas the $(a,c)$ contains fields with $(q_L,q_R)=(-1,+1)$, both
with conformal weights $(h_L,h_R)=({1\over2},{1\over2})$. The latter
are related, via spectral flow, to states in the $(c,c)$ ring with
charges $(2,1)$.

We now turn to a comparison between the chiral primary states and the
cohomology of the CY manifold. We expect a close relationship since
the (2,2) super-conformal field theories we are considering correspond
to conformally invariant sigma models with CY target space.
Let us first look at the $(c,c)$ ring. In the conformal field theory
there is a unique chiral primary state with $(q_L,q_R)=(3,0)$
whereas there is the unique holomorphic three form $\Omega$ in the
cohomology of the CY space. By  conjugation we have
the state with charge $(0,3)$ and $\bar\Omega\in H^{0,3}(X)$. Also
$(q_L,q_R)=(3,3)\leftrightarrow\Omega\wedge\bar\Omega$.
The fields with $(q_L,q_R)=({1},{1})$ are marginal and
correspond to the complex structure moduli, whereas the fields
with $(q_L,q_R)=(2,1)$ are related (via spectral flow by $(-1,0)$)
to marginal states in the $(a,c)$ ring with charges $(-1,1)$. They
correspond to the complex structure deformations.
In general, if we identify the left and right $U(1)$ charges with
the holomorphic and anti-holomorphic form degrees, respectively, one is
tempted to establish a one-to-one correspondence between elements
of the $(c,c)$ ring with charges $(q_L,q_R)$ and elements of
$H^{q_L,q_R}(X)$. This becomes even more suggestive if we formally
identify the zero modes of the
supercurrents with the holomorphic exterior differential and
co-differential as
$G_0^+\sim\p,\,\bar G_0^+\sim\bar\p$.
Via the spectral flow (by
$(\theta_L,\theta_R)=(-{1\over2},-{1\over2})$)
we can also identify
$G^+_{-{1\over2}}\sim\p$ and $\bar G^+_{-{1\over2}}\sim\bar\p$. Furthermore,
one can show \lvw~ that each $NS$ state has a chiral primary representative
in the sense that there exists a unique decomposition
$|\phi\ket=|\phi_0\ket+G^+_{-{1\over2}}|\phi_1\ket
+G_{{1\over2}}^-|\phi_2\ket$
with $|\phi_0\ket$ chiral primary. For $|\phi\ket$ itself primary,
$|\phi_2\ket$ is zero.
This parallels the Hodge decomposition of differential forms.
In fact, the one-to-one correspondence between the
cohomology of the target space of supersymmetric sigma models and
the Ramond ground states has been established in
\ref\wittensusy{E. Witten, \npb202(1982)253; R. Rohm and E Witten,
Ann. Phys. {\bf 170} (1986) 454}.
Let us now compare the $(c,c)$ ring with the cohomology ring, whose
multiplicative structure is defined by taking wedge products
of the harmonic forms. We know from \yukawas~ that in the large radius
limit the Yukawa couplings are determined by the cohomology of the
Calabi-Yau manifold $X$. However, once string effects are taken into account,
the Yukawa couplings are no longer determined by the cohomology ring of $X$
but rather by a deformed cohomology ring. This deformed cohomology ring
coincides with the $(c,c)$ ring of the corresponding super-conformal
field theory. For discussions of the relation between chiral rings
and cohomology rings in the context of topological $\sigma$-model
we refer to Witten's contribution in \mirror.

On the conformal field theory level, mirror symmetry is now the
following simple observation: the exchange of the relative
sign of the two $U(1)$ currents is a trivial symmetry of the conformal
field theory, by which the  $(c,c)$ and the $(a,c)$ rings of
chiral primary fields are exchanged. On the geometrical level this does
however have highly non-trivial implications.
It suggests the existence of two topologically very different
geometric interpretations of a given $(2,2)$ internal superconformal
field theory. The deformed cohomology rings are isomorphic to the
$(c,c)$ and $(a,c)$ rings, respectively. Here we associate elements
in the $(a,c)$ ring with charge $(q_L,q_R)$ with elements of
$H^{3+q_L,q_R}$.

If we denote the two manifolds by $X$ and $X^*$
then one simple relation is in terms of their Hodge numbers:
\eqn\duality{
h^{p,q}(X)=h^{3-p,q}( X^*)
}
The two manifolds $X$ and $X^*$ are referred to as a mirror pair.
The relation between the Hodge numbers alone is not very strong.
A farther reaching consequence of the fact that string
compactification on
$X$ and $X^*$ are identical is the relation between the
deformed cohomology rings. This in turn entails a relation between
correlation functions, in particular between the two types of
Yukawa couplings.

It is however not clear whether a given $(2,2)$ theory always allows for
two topologically distinct geometric descriptions as compactifications
\foot{Mirror symmetric manifolds are excluded if $\chi\neq0$.}.
In fact, if one considers so-called rigid manifolds, i.e. CY manifolds
with $h^{2,1}=0$, then it is clear that the mirror manifold cannot be
a CY manifold, which is K\"ahler, i.e. has $h^{1,1}\geq1$.
The concept of mirror symmetry for these cases has been exemplified
on the $\ZZ_3\times\ZZ_3$ orbifold in
\ref\cdp{P. Candelas, E. Derrick and L. Parkes, \npb407(1993)115}
and was further discussed in
\ref\schimmrigk{R. Schimmrigk, \prl70(1993)3688 and
{\sl K\"ahler manifolds with positive first Chern class and mirrors
of rigid Calabi-Yau manifolds}, preprint BONN-HE 93-47}.

To close this section we want to make some general comments about the
structure of the moduli space of Calabi-Yau compactifications which will
be useful later on (see \ferraratheisen~for a review of these issues).
In this context it is useful to note
that instead of compactifying the heterotic
string on a given Calabi-Yau manifold, one could have just as well taken the
type II string. This would result in $N=2$ space-time supersymmetry.
In the conformal field theory language this means that we take the
same $(2,2)$ super-conformal field theory with central charge
$(\bar c,c)=(9,9)$ but this time without the additional gauge sector that
was required for the heterotic string. This results in one
gravitino on the left and on the right moving side each.
The fact that the identical internal conformal field theory might also
be used to get a $N=2$ space-time supersymmetric theory leads to additional
insight into the structure of the moduli space which is the same
for the heterotic as for the type II string and which has to satisfy
additional constraints coming from the second space-time
supersymmetry. This was used in
\ref\seiberg{N. Seiberg, \npb303(1988)286}
to show that locally the moduli manifold has the product structure
\eqn\modulimanifold{
{\cal M}={\cal M}_{h^{1,1}}\times {\cal M}_{h^{2,1}}
}
where ${\cal M}_{h^{1,1}},{\cal M}_{h^{2,1}}$ are two K\"ahler manifolds
with complex dimensions $h^{1,1}$ and $h^{2,1}$, respectively.
The same result was later derived in refs.
\ref\ferraraetal{S. Cecotti, S. Ferrara and L. Girardello,
\ijmpa4(1989)2475;
S. Ferrara, Nucl. Phys. (Proc. Suppl.) 11 (1989) 342}
\ref\candelasetal{P. Candelas, P. Green and T. H\"ubsch, {\sl Connected
Calabi-Yau compactifications}, in Strings `88, J. Gates et al. (eds),
World Scientific 1989 and \npb330(1990)49;
P. Candelas, T. H\"ubsch and R. Schimmrigk, \npb329(1990)582}
\ref\dkl{L. Dixon, V. Kaplunovsky and J. Louis, \npb329(1990)27}.
In ref.\dkl~it was shown to be a consequence of the $(2,2)$ super-conformal
algebra. $N=2$ space-time supersymmetry or super-conformal Ward
identities can be used to show that each factor of ${\cal M}$
is a so-called special K\"ahler manifold. Special K\"ahler manifolds
are characterized by a prepotential $F$ from which the
K\"ahler potential (and thus the K\"ahler metric) and also the Yukawa
couplings can be computed.
Locally on special K\"ahler manifolds there exist so-called special
coordinates $t_i$ which allow for simple expressions of the
K\"ahler potential $K$ and the Yukawa couplings $\kappa$ in terms
of the prepotential as\foot{From this
it follows immediately that the $\bra\overline{27}^3\ket$ couplings receive
no instanton corrections since they would lead to a dependence on the
K\"ahler moduli.}
\ref\specialkaehler{B. de Wit, P.G. Lauwers, R. Philippe, S.Q. Su and
A. van Proeyen, \plb134(1984)37; B. de Wit and A. van Proeyen,
\npb245(1984)89; J.P. Derendinger, S. Ferrara, A. Masiero and
A. van Proeyen, \plb140(1984)307; B. de Wit, P.G. Lauwers and
A. van Proeyen, \npb255(1985)569; E. Cremmer, C. Kounnas, A. van Proeyen,
J.P. Derendinger, S. Ferrara, B. de Wit and L. Girardello,
\npb250(1985)385}
\eqn\prepotential{
\eqalign{K&=-\ln\Bigl\(2(F-\bar F)-(t_i-\bar t_i)(F_i+\bar F_i)\Bigr\)
\,,\qquad\left(F_i={\p F\over\p t_i}\right)\cr
\kappa_{ijk}&=F_{ijk}}
}
with one set for each factor of ${\cal M}$\foot{We will denote the
special coordinates on ${\cal M}_{h^{1,1}}$ by $t$ and the ones on
${\cal M}_{h^{2,1}}$ by $\lambda$.}.
To reproduce the
classical Yukawa couplings \yukawas~$F^{1,1}(X)$ is simply a cubic polynomial
whereas $F^{2,1}$ is a complicated function of the complex structure
moduli. Instanton corrections will modify $F^{1,1}$ such as to reproduce
the couplings \instantonsum.  Mirror symmetry then relates the pre-potentials
on $X$ and $X^*$. If $\lambda_i$ are the complex structure moduli on
$X^*$ then one can find a local map (the mirror map)
${\cal M}_{2,1}(X^*)\to{\cal M}_{1,1}(X):
\,\lambda_i\mapsto t_i(\lambda)$ such that
$F^{2,1}(\lambda)(X^*)=F^{1,1}(t)(X)$ and similarly for
$F^{1,1}(X^*)$  and $F^{2,1}(X)$.
We will find below that $\lambda_i(t)$ are transzendental functions
containing exponentials. It is an interesting fact that $\lambda(q^i)$,
with $q=e^{2\pi i t}$, is always an
infinite series with integer coefficients.

%%%%%%%%%%%%%%%%%%%%%%%%%%%%%%%%%%%%%%%%%%%%%%%%%%%%%%%%%%%%%%%%%%%%
\newsec{Construction of Mirror Pairs}
%%%%%%%%%%%%%%%%%%%%%%%%%%%%%%%%%%%%%%%%%%%%%%%%%%%%%%%%%%%%%%%%%%%%

We have already alluded to the fact that a classification of consistent
string vacua is still out of reach. In fact, even a classification
of three-dimensional CY manifolds, providing just a subset of string vacua,
is still lacking. (One only knows how to classify the homotopy types
of the manifolds by virtue of Wall's theorem; \huebsch, p. 173.)
What has been achieved
so far is to give complete lists of possible CY manifolds within a
given construction. But even here one does not have criteria to
decide which of these manifolds are diffeomorphic.

The constructions that have been completely searched for are
hypersurfaces in four-dimensional weighted projective space
\ref\ksks{M. Kreuzer and H. Skarke, \npb388(1993)113;
A. Klemm and R. Schimmrigk, \npb411(1994)559}
and complete intersections of $k$ hypersurfaces in $3+k$
dimensional products of projective spaces
\ref\cdls{P. Candelas, A. Dale, A. L\"utken and R. Schimmrigk,
\npb298(1988)493; P. Candelas, A. L\"utken and R. Schimmrigk,
\npb306(1989)105}.
In these notes we will limit ourselves to the discussion of
hypersurfaces.
Mirror symmetry for complete intersections has been discussed in
\ref\batyrevstraten{V. Batyrev and D. van Straten, {\sl Generalized
Hypergeometric Functions and Rational Curves on Calabi-Yau Complete
Intersections in Toric Varieties}, preprint 1992} and \hktyII.

Weighted $n$-dimensional complex projective space
$\IP^n\(\vec w\)$ is simply $\IC^{n+1}\setminus\{0\}/\IC^*$
where $\IC^*=\IC\setminus\{0\}$ acts as
$\l\cdot(z_0,\dots,z_n)=(\l^{w_0}z_0,\dots,\l^{w_n}z_n)$.
We will denote a point in $\IP^n\(\vec w\)$ by $(z_0\:\cdots\:z_n)$.
The coordinates
$z_i$ are called homogeneous coordinates of $\IP^n\(\vec w\)$
\foot{We will sometimes denote them by $z_0,\dots,z_n$ and
othertimes by $z_1,\dots,z_{n+1}$. We are confident that this will
cause no confusion.}
and $w_i\in\ZZ_+$ their weights. For $\vec w=(1,\dots,1)$ one recovers
ordinary projective space. Note however that $\IP^n\(k\vec w\)\sim
\IP^n\(\vec w\)$. In fact, due to this and other isomorphisms
(see Prop. 1.3.1 in
\ref\dolgachev{I. Dolgachev, {\sl Weighted Projective Varieties} in
{\sl Lecture Notes in Mathematics} 956 Springer-Verlag (1992) 36})
one only needs to consider so-called
well-formed weighted projective spaces. $\IP^n\(\vec w\)$ is called
well-formed if each set of $n$ weights is co-prime.
Weighted projective $\IP^n\(\vec w\)$ can be covered with $n+1$ coordinate
patches $U_i$ with $z_i\neq0$ in $U_i$.
The transition functions between different patches are then
easily obtained. The characteristic feature to note about
the transition functions of projective space is that
in overlaps $U_i\cap U_j$ they are given by Laurent monomials. For example,
consider $\IP^2$ with homogeneous coordinates $(z_0\:z_1\:z_2)$ and the
three patches $U_i$, $i=0,1,2$ with inhomogeneous coordinates
$\varphi_0(z_0\:z_1\:z_2)=(u_0,v_0)
=({z_1\over z_0},{z_2\over z_0})$,
$\varphi_1(z_0\:z_1\:z_2)=(u_1,v_1)
=({z_0\over z_1},{z_2\over z_1})$ and
$\varphi_2(z_0\:z_1\:z_2)=(u_2,v_2)
=({z_0\over z_2},{z_1\over z_2})$.
The transition functions on overlaps are then Laurent monomials;
e.g. on $U_0\cap U_1$ we have $(u_1,v_1)=({1\over u_0},
{v_0\over u_0})$, and we have $\varphi_i(U_i\cap U_j)
\simeq\IC\times \IC^*,\,\varphi_i(U_1\cap U_2\cap U_3)=(\IC^*)^2$.
The reason for dwelling on these well-known facts is
that below we will treat this example also
in the language of toric geometry and that it is the fact that
transition functions between patches are Laurent monomials,
that characterizes more general toric varieties.

Weighted projective spaces are generally singular. As an example, consider
$\IP^2\(1,1,2\)$, i.e. $(z_0,z_1,z_2)$ and $(\l z_0,\l z_1,\l^2 z_2)$
denote the same point and for $\l=-1$ the point
$(0\:0\:z_2)\equiv(0\:0\:1)$
is fixed but $\l$ acts non-trivially on its neighborhood:
we thus have a $\ZZ_2$ orbifold singularity at this point.

A hypersurface $X$ in (weighted) projective space is defined as the vanishing
locus of a (quasi)homogeneous polynomial, i.e. of a polynomial
in the homogeneous coordinates that satisfies
$p(\l^{w_0}z_0,\dots,\l^{w_n}z_n)=\l^d p(z_0,\dots,z_n)$ where
$d\in\ZZ_+$ is called the degree of $p(z)$; i.e. we have
\eqn\hypersurface{
X_{\vec w}=\left\{(z_0\:\dots\:z_n)\in\IP^n\(\vec w\)\Big|p(z)=0\right\}
}
In order for the hypersurface to be a CY manifold one has to require that
its first Chern class vanishes. This can be expressed in terms of the
weights and the degree of the defining polynomial as\huebsch
\eqn\chernone{
c_1(X)=\Bigl(\sum_{i=0}^n w_i-d\Bigr)J
}
where $J$ is the K\"ahler form of the projective space the manifold
is embedded in. A necessary condition for a hypersurface
in $\IP^4\(\vec w\)$ to be a three-dimensional CY manifold
is then that the degree of the defining polynomial equals the sum
of the weights of the projective space. However one still has to
demand that the embedding $X\hookrightarrow\IP^n\(\vec w\)$
be smooth. This means that
one has to require the transversality condition: $p(z)=0$ and $dp(z)=0$
have no simultaneous solutions other than $z_0=\dots=z_n=0$
(which is not a point of $\IP^n\(\vec w\)$).

There exists an easy to apply criterium
\ref\fletcher{A. R. Fletcher,
{\sl Working with Weighted Complete Intersections},
Max-Planck-Institut Series, No. 35 (1989) Bonn}
which allows one to decide whether a given polynomial satisfies
the transversality condition. This criterium follows from Bertini's theorem
(c.f. e.g. Griffiths and Harris in \cy) and goes as follows.
For every index set $J=\{j_1,\ldots,j_{|J|}\} \subset
\{0,\ldots,n\}=N$ denote by $z^{m^{(k)}}_J$ monomials
$z_{j_1}^{m_{i_1}^{(k)} }\cdots z_{i_{|J|}}^{m^{(k)}_{i_{|J|}}}$ of
degree $d$.
Transversality is then equivalent to the condition that for every
index set $J$ there exists either $(a)$ a monomial $z_J^m$, or
$(b)$ $|J|$-monomials $z_J^{m^{(k)}}z_k$ with $|J|$ distinct
$k\in N\setminus J$.

Analysis of this condition shows that there are 7555 projective
spaces $\IP^4\(\vec w\)$ which admit transverse hypersurfaces.
They were classified in \ksks. In case that $q_i=d/w_i\in\ZZ,\,
\forall i=0,\dots,k$ one gets a transverse polynomial of Fermat type:
$p(z)=\sum_{i=0}^4 z_i^{q_i}$.

If the hypersurfaces $X$ meet some of the singularities
of the weighted projective space, they are themselves singular.
Let us first see what kind of singular sets one can get.
If the weights $w_i$ for $i\in I$ have a common factor $N_I$,
the singular locus $S_I$ of the CY space is the intersection
of the hyperplane $H_I=\{z\in\IP^4\(\vec w\)|z_i=0
\,{\rm for}\, i \notin\, I\}$ with $X_{\vec w}$.
We will see that the singular locus consists either of points or of
curves.

As $\IP^4[\vec w]$ is wellformed we have $|I|\le 3$.
Consider $|I|=3$ and apply the transversality criterum to $J=I$. Obviously,
only transversality condition $(a)$ can hold, which implies that $p$
will not vanish identically on $H_I$, hence  $\dim(S_I)=1$.
It is important for the following to consider the $\IC^*$-action
on the normal bundle\foot{The normal bundle on $C$ is the
quotient bundle $N_C=T_X|_C/T_C$.}
to this curve. We write the $c_1(X)=0$ condition
as $\sum_{i\in I} w'_i + \sum_{j\notin I} (w_j/N_I)=(d/N_I)$, with
$w'_i\in \ZZ_>$ and $(w_j/N_I)\notin\ZZ$. Because of $(a)$ one has
$d=\sum_{i\in I} m_i w_i=N_I\sum_{i\in I} m_i w'_i $, for
$m_i\in \ZZ_\ge$, from which we can conclude
$\sum_{j\notin I} (w_i/N_I)\in\ZZ$.
Locally we can then choose $(z_{k_1},z_{k_2})$ with $k_i\notin I$
as the coordinates normal to the curve. The $\IC^*$-action
which fixes $S_I$ will therefore be generated by
$(z_{k_1},z_{k_2})\mapsto (\lambda z_{k_1},\lambda^{-1}z_{k_2})$, where
we define $\l=e^{2\pi i/N_I}$. That is, locally the singularity
in the normal bundle is of type $\IC^2/\ZZ_{N_I}$.

Finally for $|I|\le 2$ clearly $\dim (S_I)\le 1$. From the analysis of
the divisibility condition imposed by transversality and $c_1(X)=0$
we can summarize that the $\IC^*$ action in the normal bundle of
singular curves or the neighborhood of singular
points is in local coordinates
always of the form $\IC^2/\ZZ_{N_I}$ and $\IC^3/\ZZ_{N_I}$ with
\eqn\cstar{
\eqalign{\l\cdot(z_1,z_2)&=(\l z_1,\l^{-1}z_2)\cr
\l\cdot(z_1,z_2,z_3)&=(\l z_1,\l^a z_2,\l^b z_3)
         \quad{\rm with}\quad 1+a+b=N_I,\,a,b\in \ZZ}}
Note that for the case of fixed curves invariant monomials are
$z_1^{N_I},z_2^{N_I}$ and $z_1 z_2$, i.e. we can describe the
singularity as $\{(u,v,w)\in\IC^3|uv=w^{N_I}\}$. This type of singularity is
called a rational double point of type $A_{N_I-1}$. The relation with the
Lie-algebras from the $A$ series and the discussion
of the resolution of the singularities within toric
geometry will be given below.

Let us mention that the types of singularities encountered here
are the same as the ones in abelian toroidal orbifolds, discussed in
\ref\mop{D. G. Markushevich, M. A. Olshanetsky and A. M. Perelomov,
\cmp111(1987)247}
\ref\ek{J. Erler and A. Klemm, Comm. Math. Phys. 153 (1993) 57}.
The condition on the exponents of the $\IC^*$-action there is
related to the fact that one considers only subgroups of
$SU(3)$ as orbifold groups. This was identified as a necessary
condition to project out of the spectrum
three gravitinos
in order to obtain exactly $N=1$ spacetime supersymmety
\ref\orb{L. Dixon, J. Harvey, C. Vafa and E. Witten,
Nucl. Phys. B261 (1985) 678 and Nucl. Phys. B274 (1986) 285}.
As we will briefly explain below, it also ensures $c_1(\hat X)=0$,
i.e. triviality of the canonical bundle of the resolved manifold $\hat X$.
Desingularisations with this property are referred to as minimal
desingularizations
\ref\ry{S.S. Roan and S. T. Yau, Acta Math. Sinica (NS) 3 (1987) 256;
S.S. Roan, J. Diff. Geom. 30 (1989) 523}.
In ref.\ksks~ the Hodge numbers of the minimal desingularizations
of all 7555 transverse hypersurfaces were evaluated.

Among the singular points we have to distinguish between isolated points
and exceptional points, the latter being singular points on singular
curves or the points of intersection of singular
curves. The order $N_I$ of the isotropy group $I$ of exceptional
points exceeds that of the curve.
In order to get a smooth CY manifold $\hat X$,
these singularities have to be resolved
by removing the singular sets and replacing them by smooth
two complex dimensional manifolds which are then called exceptional divisors.
Each exceptional divisor $D$ provides, by Poincar\'e duality, a harmonic
$(1,1)$ form $h_D$: $\int_D\a=\int_{\hat X} \a\wedge h_D$ for
every closed $(2,2)$ form $\a$ and $h^{1,1}(\hat X)=\#$ exceptional
divisors $+1$ where the last contribution counts the restriction of the
K\"ahler form from to embedding space to $\hat X$ (to which one can also
associate a divisor).

There are only few hypersurfaces, namely the quintic
(i.e. $d=5$) in $\IP^4$, the sextic in $\IP^4\(1,1,1,1,2\)$,
the octic in $\IP^4\(1,1,1,1,4\)$
and finally the dectic in $\IP^4\(1,1,1,2,5\)$, which do not require
resolution of singularities and the K\"ahler form they inherit from the
embedding space is in fact the only one, i.e. for these CY spaces
$h^{1,1}=1$. One easily sees that these Fermat hypersurfaces do not
meet the singular points of their respective embedding spaces.
They were analyzed in view of mirror symmetry in \morrison.

If one considers the lists of models of ref.\ksks~ one finds already
on the level of Hodge numbers that there is no complete mirror symmetry
within this construction. Also, if one includes abelian orbifolds
of the hypersurfaces
\ref\ksn{M. Kreuzer and H. Skarke, Nucl. Phys. B405 (1993) 305;
         M. Kreuzer, Phys. Lett. 314B (1993) 31;
         A. Niemeyer, Diplom Thesis, TU-M\"unchen, 1993}
the situation does not improve.

One can however get a mirror symmetric set of CY manifolds if one
generalizes the construction to hypersurfaces in so-called
toric varieties. Since they are not (yet) familiar to most physicists
but relevant for describing the resolution of the
above encountered singularities and for the discussion of mirror symmetry,
we will give a brief description of toric varieties. For details and
proofs we have to refer to the literature
\ref\toric{T. Oda, {\sl Convex Bodies and Algebraic Geometry:
an Introduction to the Theory of Toric Varieties}, Ergebnisse der
Mathematik und ihrer Grenzgebiete, 3. Folge, Bd. 15, Springer Verlag
1988; V.I. Danilov, Russian Math. Surveys, 33 (1978) 97;
M. Audin, {\sl The Topology of Torus Actions on Symplectic
Manifolds},
Progress in Mathematics, Birkh\"auser 1991;
W. Fulton, {\sl Introduction to Toric Varieties}, Annals of Mathematics
Studies, Princeton University Press 1993;
V.I. Arnold, S. M. Gusein-Zade and A.N. Varchenko, {\sl Singularities of
Differntial Maps}, Birkh\"auser 1985, Vol. II, Chapter 8}.
Toric methods were first used in the
construction of Calabi-Yau manifolds in \ry~and \mop.
They have entered the discussion of mirror symmetry through
the work of V. Batyrev
\ref\batyrevI{V. Batyrev, Duke Math. Journal 69 (1993) 349}
\ref\batyrevII{V. Batyrev, Journal Alg. Geom., to be published}
\ref\batyrevIII{V. Batyrev, {\sl Quantum Cohomology Rings of Toric
Manifolds}, preprint 1992}
\batyrevstraten
\ref\batyrevcox{V. Batyrev and D. Cox, {\sl On the Hodge Structure of
Projective Hypersurfaces in Toric Varieties}, preprint 1993}.

Toric varieties are defined in terms of a lattice $N\simeq\ZZ^n$
and a fan $\Sigma$. Before explaining what a fan is, we first have to
define a ({\it strongly convex rational polyhedral}) cone
(or simply cone) $\sigma$ in the real vector space
$N_\IR\equiv N\otimes_{\ZZ}\IR$:
\eqn\cone{
\sigma=\Bigl\{\sum_{i=1}^s a_i n_i; a_i\geq0\Bigr\}
}
where $n_i$ is a finite set of {\it lattice} vectors
(hence rational), the generators of the cone. We often write simply
$\bra n_1,\dots,n_s\ket$. Strong convexity
means that $\sigma\cap(-\sigma)=\{0\}$, i.e. the cone does not
contain lines through the origin.
A cone $\sigma$ is called {\it simplicial} if it is generated
by linearly independent (over $\IR$) lattice vectors. A cone generated
by part of a {\it basis} of the lattice
$N$ is called a basic cone. If we normalize
the unit cell of the lattice to have volume one, then a simplicial cone
$\bra n_1,\dots n_m\ket$ is basic if $\det(n_1,\dots,n_m)=1$ (here
$n_i$ are the generators of minimal length).
If a cone fails to be basic, not all lattice
points within the cone can be reached
as linear combinations of the generators with positive integer
coefficients.

To every cone we can define the dual cone $\sigma^\vee$ as
\eqn\dualcone{
\sigma^\vee=\Bigl\{x\in M_\IR;\bra x,y\ket\geq0\,\,
{\rm for~all}\,\,y\in\sigma\Bigr\}
}
where $M$ is the lattice dual to $N$ and $\bra\,,\ket:M\times N\to\ZZ$,
which extends to the $\IR$ bilinear pairing $M_\IR\times N_\IR\to\IR$.
$\sigma^\vee$ is rational with respect to
$M$ but strongly convex only if $\dim(\sigma)=\dim (M_{\IR})$.
(For instance the cones dual to one-dimensional
cones in $\IR^2$, are half-planes.). Also
$(\sigma^\vee)^\vee=\sigma$. Given a cone we define
\eqn\semigroup{S_\sigma=\sigma^\vee\cap M
}
which is finitely generated by say $p\geq{\rm dim}\,M\equiv d_M$
lattice vectors $n_i$.
In general (namely for $p\geq{d_M}$) there will be non-trivial
linear relations between the generators of $S_\sigma$ which can be written
in the form
\eqn\toricrelations{
\sum\mu_i n_i=\sum\nu_i n_i
}
with $\mu_i$ and $\nu_i$ non-negative integers.
A cone then defines an affine toric variety as
\eqn\Usigma{
U_\sigma=\{(Z_1,\dots,Z_p)\in\IC^p|Z^\mu=Z^\nu\}
}
where we have used the short hand notation
$Z^\mu=Z_1^{\mu_1}\cdots Z_p^{\mu_{d_M}}$.
In the mathematics literature one often finds the notation
$U_\sigma={\rm Spec}(\IC\(S_\sigma\))={\rm
Spec}(\IC\(Z_1,\dots,Z_p\)/I)$
where the ideal $I$ is generated by all the relations $Z^\mu=Z^\nu$
between the gerators of $S_\sigma$.

To illustrate the construction, let us look at a simple example.
Consider the cone $\sigma$ generated by $(N+1)e_1-N e_2$ and $e_2$.
Then $S_\sigma$ is easily shown to be generated by $n_1=e_1^*,\,
n_2=N e_1^*+(N+1)e_2^*$ and $n_3=e_1^*+e_2^*$ which satisfy
$(N+1)n_3=n_1+n_2$. This leads to
\eqn\Usigmaexample{
U_\sigma=\{(Z_1,Z_2,Z_3)\in\IC^3|Z_1 Z_2=Z_3^{N+1}\}
}
This is just the $A_{N}$ rational double point discussed above.

A face $\tau$ of a cone $\sigma$ is what one expects;
it can be defined as
$\sigma\cap u^\perp=\{v\in \sigma:\bra u,v\ket=0\}$ for
some $u\in\sigma^\vee$. The constructive power of toric
geometry relies in the possibility of gluing cones to fans.

A fan $\Sigma$ is a family of cones $\sigma$ satisfying:
\item{$(i)$} any face of a cone in $\Sigma$ is itself a cone in
$\Sigma$;
\item{$(ii)$} the intersection of any two cones in $\Sigma$ is a face
              of each of them.

After having associated an affine toric variety $U_\sigma$ with a cone
$\sigma$, we can now construct a toric variety $\IP_\Sigma$ associated
to a fan $\Sigma$ by glueing together the
$U_\sigma,\,\sigma\in\Sigma$:
\eqn\XSigma{
\IP_\Sigma=\bigcup_{\sigma\in\Sigma} U_\sigma
}
The $U_\sigma$ are open subsets of $\IP_\Sigma$. The glueing works because
$U_{\sigma\cap\sigma'}$ is an open subset of both $U_\sigma$ and $U_\sigma'$
i.e. $U_{\sigma \cap \sigma'}=U_\sigma\cap U_{\sigma'}$.
On the other hand $U_{\sigma^\vee \cap (\sigma')^\vee}\ne
U_{\sigma^\vee}\cap U_{(\sigma')^\vee}$,
hence gluing is natural for the cones $\sigma$.
A last result we want to quote before demonstrating the above
with an example is that a toric variety $\IP_\Sigma$ is compact iff the union
of all its cones is the whole space $\IR^n$. Such a fan is called complete.

To see what is going on we now give a simple but representative example.
Consider the fan $\Sigma$ whose dimension one cones
$\tau_i$ are generated by the
vectors $n_1=e_1,\,n_2=e_2$ and $n_3=-(e_1+e_2)$. $\Sigma$ also contains
the dimension two cones $\sigma_1:\bra n_1,n_2\ket,\,\sigma_2:\bra n_2,n_3\ket
,\,\sigma_3:\bra n_3,n_1\ket$ and of course the dimension zero cone $\{0\}$.
We first note that $\Sigma$ satisfies the compactness criterium.
The two-dimensional dual cones are
$\sigma_1^\vee:\bra e_1^*,e_2^*\ket,\,
\sigma_2^\vee:\bra-e_1^*,-e_1^*+e_2^*\ket,\,
\sigma_3^\vee:\bra-e_2^*,e_1^*-e_2^*\ket$ and
$U_{\sigma_1}={\rm Spec}(\IC\(X,Y\)),\,
U_{\sigma_2}={\rm Spec}(\IC\(X^{-1},X^{-1}Y\)),\,
U_{\sigma_3}={\rm Spec}(\IC\(Y^{-1},XY^{-1}\))$
each isomorphic to $\IC^2$. These glue together to form
$\IP^2$. Indeed, if we define coordinates $u_i,v_i$ via
$U_{\sigma_i}={\rm Spec}(\IC\(u_i,v_i\))$ we get the transition functions
for $\IP^2$ between the three patches. Note also that
$U_{\tau_i}=U_{\sigma_i\cap\sigma_j}\simeq \IC\times\IC^*$
and $U_{\sigma_1\cap\sigma_2\cap\sigma_3}=U_{\{0\}}=(\IC^*)^2$.
To get e.g. the weighted projective space $\IP^2\(1,2,3\)$ one simply
has to replace the generator $n_3$ by $n_3=-(2 e_1+3 e_2)$. In fact,
all projective spaces are toric varieties. This will become clear below.

We now turn to the discussion of singularities of toric varieties and
their resolution. We have already given the toric description of
the rational double point. The reason why the corresponding cone leads
to a singular variety is because it is not basic, i.e.
it can not be generated by a basis of the lattice $N$. This in
turn results in the need for three generators for $S_\sigma$ which
satisfy one linear relation. The general statement is that
$U_\sigma$ is smooth if and only if $\sigma$ is basic.
Such a cone will also be called smooth. An $n$-dimensional toric variety
$X_\Sigma$ is smooth, i.e. a complex manifold, if and only if
all dimension $n$ cones in $\Sigma$ are smooth. We do however have to require
more than smoothness. We also want to end up with a CY manifold, i.e.
a smooth manifold with $c_1=0$.

It can be shown that if $\Sigma$ is a smooth fan (i.e. all its cones
are smooth) $X_\Sigma$ has trivial canonical bundle if and only if the
endpoints of the minimal generators of
all one-dimensional cones in $\Sigma$ lie on a hyperplane
(see the proof in the appendix to \mop).
The intersection of the hyperplane with the fan is
called the trace of the fan.
%Note that
%this was the case for the resolution of the rational double point (and will
%also be satisfied by the resolution of the point singularities
%discussed below).
An immediate consequence of this result is that
a compact toric variety, i.e. corresponding to a complete fan, can
never have $c_1=0$. One therefore has to consider hypersurfaces in compact
toric varieties to obtain CY manifolds.

The singularities we are interested in are cyclic quotient singularities.
A standard result in toric geometry states that $X_\Sigma$ has only
quotient singularities, i.e. is an orbifold,
if $\Sigma$ is a simplicial fan, i.e. if all cones
in $\Sigma$ are simplicial.
Given a singular cone one resolves the singularities by subdividing the
cone into a fan such that each cone in the fan is basic.

Let us demonstrate this on the rational $A_{N}$ double point.
The two-dimensional cone $\sigma^{(2)}$ was generated
by $(N+1)e_1-Ne_2$ and $e_2$ which is not a basis of the lattice
$N=\ZZ^2$. We now add the one-dimensional cones $\sigma^{(1)}_m$
generated by
$m e_1-(m-1)e_2$ for $m=1,\dots,N$. (In this notation $\sigma_0$ and
$\sigma_{N+1}$ are the original one-dimensional cones.) The
two-dimensional cones $\sigma^{(2)}_m:\bra m e_1-(m-1)e_2,(m+1)e_1-m e_2\ket,
\,m=1,\dots,N$ are then basic and furthermore we have a $c_1=0$ resolution.

The original singular manifold has thus been desingularized by gluing
exceptional divisors $D_m,\,m=1,\dots,N$.
To cover the nonsingular manifold
one needs $N+1$ patches. One generator of
$(\sigma_i^{(2)})^\vee$ and one of
$(\sigma_{i+1}^{(2)})^\vee$ are antiparallel in
$\Sigma^\vee$. These generators therefore correspond
to inhomogenous coordinates of one of $N$ $\IP^1$'s.
The exceptional divisors are therefore $\IP^1$'s.
By inspection of the various patches one can see
that the  $\IP^1$'s intersect pairwise transversely in
one point to form a chain.
The self-intersection number is obtained
as the degree of their normal bundles,
which is $-2$, so that the intersection matrix
is the (negative of) the Cartan matrix of $A_N$;
for details we refer to Fulton, ref.\toric.
Such a collection of $\IP^1$'s is called a
Hirzebruch-Jung sphere-tree
\ref\Hirzebruch{F. Hirzebruch, Math. Annalen 124 (1951) 77}.
Recalling that the rational double points appeared in
the discussion of {\it curve} singularities of CY
manifolds we have seen that the resolved singular curves are locally the
product of the curve $C$ and a Hirzebruch-Jung sphere-tree.

Data of toric varieties depend only on linear relations and
we may apply bijective linear transformations to
choose a convenient shape. E.g. given the canonical
basis $e_1$, $e_2$ in $\IR^2$, we may use $e_i$
as generators of the cone and
$n_1={1\over N+1}e_1+{N\over N+1}e_2$, $n_2=e_2$, as generators
of $N=\IR^2$ to describe the $A_N$ singularity.
This form generalizes easily to higher dimensional cyclic
singularities such as
$\IC^3/\ZZ_N$, the general form of the
{\it point} singularities \cstar. They are described
by $e_i$, $i=1,2,3$ as generators of the cone and the lattice basis
$n_1={1\over N}(e_1+a e_2+b e_3),\,n_2=e_2,\,n_3=e_3$.
The local desingularisation process consists again
of adding further generators such as to obtain a smooth fan.
One readily sees that as a consequence of \cstar~all
endpoints of the vectors generating the nonsingular fan
lie on the plane $\sum_{i=1}^3 x_i=1$, as it is
necessary for having a trivial canonical
bundle on the resolved manifold.
The exceptional divisors are in 1-1 correspondence
with lattice points inside the cone on this hyperplane (trace of the
fan).
Their location is given by
\eqn\trace{
{\cal P}=\left\lbrace\sum_{i=1}^3 \vec e_i {a_i\over N}\left|
(a_1,a_2,a_3)\in\ZZ^3,
\pmatrix{e^{2\pi i {a_1\over N}}&&\cr&e^{2\pi i {a_2\over N}}
&\cr&&e^{2\pi i {a_3\over N} }\cr}\right.
\in \ZZ_{N},\sum_{i=1}^3 a_i=N,a_i\geq 0\right\rbrace
}
The generator  of the isotropy
group $\ZZ_{N}$ on the coordinates of the normal bundle of
the singular point is given in \cstar~and
$\vec e_1,\vec e_2,\vec e_3$ span an equilateral triangle
{}from its center. The corresponding divisors can
all be described by compact toric surfaces which have been classified.

The toric diagrams for the resolution of singular points are thus
equilateral triangles with interior points. If we have exceptional points then
there will also be points on the edges, which represent the traces
of the fans of the singular curves.

Whereas in the case of curve singularities, whose resolution was
unique, this is not so for point singularities. Given the trace with
lattice points in its interior, there are in general several ways to
triangulate it. Each way corresponds to a different resolution.
They all lead to topologically different smooth manifolds
with the same Hodge numbers,
differing however in their intersection numbers.

To obtain a K\"ahler manifold one also has to ensure
that one can construct a positive K\"ahlerform \kaehlerconditions.
%or equivalently an ample line bundle.
This is guaranteed if one can
construct an upper convex piecewise linear function on
$\Sigma$ (see Fulton, ref. \toric).
%Because of the
%Kodaira embedding theorem (see e.g. Griffiths and Harris in \cy)
%this implies that the manifold is projective algebraic, i.e.
%it can be described by algebraic constraints in some $\IP^n$.
The statement is then (see e.g. \batyrevIII)
that the cone $K(\Sigma)$ consisting of the
{\it classes} of all upper convex piecewise linear functions
(i.e. modulo globally linear functions) on $\Sigma$ is isomorphic to the
K\"ahler cone of $\IP_\Sigma$. This construction is reviewed in \hktyI.
Below we will consider hypersurfaces in $\IP_\Sigma $. The positive
K\"ahlerform on the hypersurface is then the one induced from the
K\"ahler form on $\IP_\Sigma$.

We will now give another description of toric varieties, namely in terms
of convex integral
polyhedra. The relation between the two constructions will
become clear once we have demonstrated how to extract the cones and
the fan from a given polyhedron. The reason for introducing them
is that this will allow for a simple and appealing description of
the mirror operation and of the construction of mirror pairs \batyrevII.

We start with a few definitions. We consider {\it rational} (with respect to a
lattice $N$) {\it convex polyhedra} (or simply polyhedra)
$\Delta\subset N_\IR$ containing the origin $\nu_0=(0,0,0,0)$.
$\Delta$ is called {\it reflexive} if its dual defined by
\eqn\dual{
\Delta^*=\{\;(x_1,\dots,x_4) \;\vert\;
\sum_{i=1}^4 x_i y_i \geq -1 \;{\rm for \; all \; }
(y_1,\dots,y_4)\in\Delta\;\}
}
is again a rational polyhedron. Note that if $\Delta$ is reflexive,
then $\Delta^*$ is also reflexive since $(\Delta^*)^*=\Delta$.
We associate to $\Delta$ a complete rational fan $\Sigma(\Delta)$
whose cones are the cones over the faces of $\Delta$ with apex at
$\nu_0$.

The $l$ dimensional faces will be denoted by $\Theta_l$. Completeness
is ensured since $\Delta$ contains the origin. The toric variety
$\IP_\Delta$ is then the toric variety associated to the fan
$\Sigma(\Delta^*)$,
i.e. $\IP_\Delta\equiv\IP_{\Sigma(\Delta^*)}$.

Denote by $\nu_i,\,i=0, \dots,s$ the integral points in $\Delta$ and
consider the affine space $\IC^{s+1}$ with coordinates $a_i$.
We will consider the zero locus $Z_f$ of the Laurent polynomial
\eqn\laurentpolynomial{
f_\Delta(a,X) = a_0-\sum_i a_i X^{\nu_i},\quad f_\Delta(a,X)
\in \IC[X_1^{\pm 1},\dots,X_{4}^{\pm 1}]}
in the algebraic torus $(\IC^*)^4 \subset \IP_\Delta$,
and its closure $\bar Z_f$ in $\IP_\Delta$. Here we have used the
convention $X^\mu\equiv X_1^{\mu_1}\ldots X_{4}^{\mu_{4}}$.
%The other way round, given a Laurent polynomial $f$ one defines the
%Newton polyhedron $\Delta(f)$
%as the convex hull in $N_\IR$ of all elements $\nu_i\in N$
%such that $a_i\neq 0$. Every Laurent polynomial then defines
%an affine hypersurface
%$Z_f=\{(X-1,\dots,X_4)\in (\IC^*)^*|f_{\Delta(f)}(a,X)=0\}$.
Note that by rescaling the four coordinates $X_i$ and adjusting an
overall normalization we can set five of the parameters $a_i$ to one.

$f\equiv f_\Delta$ and $Z_f$ are called $\Delta$-regular if
for all $l=1,\dots,4$
the $f_{\Theta_l}$ and $X_i{\p\over\p X_i}f_{\Theta_l},\,
\forall i=1,\dots,n$ do not vanish simultaneously in
$(\IC^*)^4$. This is equivalent to the transversality condition
for the quasi-homogeneous polynomials $p$.
Varying the parameters $a_i$ under the condition of
$\Delta$-regularity, we get a family of toric varieties.

In analogy with the situation of hypersurfaces in $\IP^4\(\vec w\)$, the more
general ambient spaces $\IP_{\Delta}$
and so $\bar Z_{f}$ are in general singular.
$\Delta$-regularity ensures that the only singularities of $\bar Z_{f}$
are the ones inherited from the ambient space.
$\bar Z_{f}$ can be resolved to a CY manifold $\hat Z_f$ iff $\IP_{\Delta}$
has only so-called Gorenstein singularities, which is the case
iff $\Delta$ is reflexive \batyrevII.

The families of  the CY manifolds $\hat Z_f$ will be denoted
by ${\cal F}(\Delta)$. The above definitions proceed in an exactly
symmetric way for the dual polyhedron $\Delta^*$ with its integral points
$\nu_i^*\;(i=0,\cdots,s^*)$, leading to families of CY manifolds
${\cal F}(\Delta^*)$.

In ref. \batyrevII~ Batyrev observed that a pair of reflexive
polyhedra
$(\Delta, \Delta^*)$ naturally provides a pair of mirror
CY families $({\cal F}(\Delta),{\cal F}(\Delta^*))$ as
the following identities for the Hodge numbers hold
\eqn\hodgenumbers{
\eqalign{h^{1,1}(\hat Z_{f,\Delta^*})&=h^{2,1}(\hat Z_{f,\Delta})\cr
&=l(\Delta)-5-\sum_{{\rm codim}\Theta=1}
l^\prime(\Theta)
+\sum_{{\rm codim}\Theta=2 }
l^\prime(\Theta)l^\prime(\Theta^*)\cr
h^{1,1}(\hat Z_{f,\Delta})&=h^{2,1}(\hat Z_{f,\Delta^*})\cr
&=l(\Delta^*\,)-5-\sum_{{\rm codim}\Theta^*=1}l^\prime(\Theta^*)
+\sum_{{\rm codim}\Theta^*=2}l^\prime(\Theta^*)
l^\prime(\Theta).}
}

Here $l(\Theta)$ and $l^\prime(\Theta)$ are the number of integral
points on a face $\Theta$ of $\Delta$ and in its interior, respectively
(and similarly for $\Theta^*$ and $\Delta^*$).
An $l$-dimensional face $\Theta$ can be represented by specifying
its vertices
${\rm v}_{i_1},\cdots,{\rm v}_{i_{k}}$. Then the dual face defined by
$\Theta^*=\{ x\in \Delta^*\; \vert \; (x,{\rm v}_{i_1})=\cdots=(x,{\rm
v}_{i_k})=-1 \}$ is a $(n-l-1)$-dimensional face of $\Delta^*$.
By construction $(\Theta^*)^*=\Theta$, and we thus have a
natural pairing between $l$-dimensional faces of $\Delta$ and
$(n-l-1)$-dimensional
faces of $\Delta^*$.  The last sum in each of the
two equations in (2.4) is over pairs of dual faces.
Their contribution cannot be associated with a monomial
in the Laurent polynomial\foot{In the language of
Landau-Ginzburg theories, if appropriate, they correspond to
contributions from twisted sectors.}. We will denote by
$\tilde h^{2,1}$ and $\tilde h^{1,1}$ the expressions
\hodgenumbers~ without the last terms.

A sufficient criterion for the possibility to
associate to a CY hypersurface in $\IP^4\(\vec w\)$ a reflexive
polyhedron is
that $\IP^4\(\vec w\)$ is Gorenstein,
which is the case if  ${\rm lcm}[w_1,\ldots,w_{5}]$
divides the degree $d$
\ref\belrtametti{M. Beltrametti and L. Robbiano,
Expo. Math. 4 (1986) 11.}.
In this case we can define a simplicial, reflexive polyhedron
$\Delta(\vec w)$ in terms of the weights, s.t. $\IP_{\Delta^*(\vec w)}
\simeq\IP^4\(\vec w\)$.
The associated $n$-dimensional integral convex dual
polyhedron is the convex hull of the integral vectors $\mu$ of the exponents
of all quasi-homogeneous monomials $z^\mu$ of degree $d$,
shifted by $(-1,\ldots,-1)$:
\eqn\polyhedron{
\Delta^*(\vec w):=
\{(x_1,\ldots, x_{5}) \in
\IR^{5}|\sum_{i=1}^{5} w_i x_i=0,x_i\geq-1\}.}
Note that this implies that the origin is the only
point in the interior of $\Delta$.

If the quasihomogeneous polynomial $p$ is Fermat, i.e. if it consists of
monomials $z_i^{d/w_i}\,\,(i=1,\cdots,5)$,
$\IP^4\(\vec w\)$ is clearly Gorenstein,
and $(\Delta,\Delta^*)$ are thus simplicial.
If furthermore at least one weight
is one (say $w_5=1$) we may choose
$e_1=(1, 0, 0, 0, -w_1)$,  $e_2 = (0, 1, 0, 0,-w_2)$,
$e_3=(0, 0, 1, 0, -w_3)$ and $e_4 = (0, 0, 0, 1,-w_4)$
as generators for $\Lambda$, the lattice induced
{}from the $\ZZ^{5}$ cubic
lattice on the hyperplane $H=\{(x_1,\ldots, x_{5}) \in
\IR^{5}|\sum_{i=1}^{5} w_i x_i=0\}$.
For this type of models we then always obtain as
vertices of $\Delta^*(\vec w)$ (with respect to the basis $e_1,\dots,e_4$)
\eqn\verticesD{
\eqalign{
&\nu_1^*=(d/w_1-1, -1, -1, -1)  \; ,\; \cr
&\nu_4^*=(-1, -1, -1, d/w_4-1)   \; ,\;  \cr }
\eqalign{
&\nu_2^*=(-1, d/w_2-1, -1, -1)  \; ,\; \cr
&\nu_5^*=(-1,-1,-1,-1)                 \;  \cr}
\eqalign{
&\nu_3^*=(-1, -1, d/w_3-1, -1)  ,\cr
& \cr}
}
and for the vertices of the dual simplex $\Delta(w)$ one finds
\eqn\vertices{
\eqalign{
&\nu_1=(1,0,0,0)  \; ,\;
 \nu_2=(0,1,0,0)  \; ,\;
 \nu_3=(0,0,1,0)  \; ,\;
 \nu_4=(0,0,0,1)  \cr
&\nu_5=(-w_1,-w_2,-w_3,-w_4)}.
}

It should be clear from our description of toric geometry that the
lattice points in the interior of
faces of $\Delta$ of dimensions $4>d>0$ correspond to exceptional
divisors resulting from the resolution of the Gorenstein
singularities of
$\IP_{\Delta^*}$. This in turn means that the corresponding Laurent monomials
in $f_\Delta$ correspond to exceptional divisors.
For those CY hypersurfaces that can be written
as a quasi-homogeneous polynomial constraint in $\IP^4\(\vec w\)$,
we can then give a correspondence between the monomials, which correspond
(via Kodaira-Spencer deformation theory) to the complex
structure moduli, and the divisors, which correspond to the
K\"ahler moduli. The authors of ref.
\ref\topologychange{D.Aspinwall, B.Greene and D.Morrison,
Phys.Lett. 303B (1993) 249 and
{\sl The Monomial-Divisor Mirror Map}, preprint IASSNS-HEP-93/43}
call this the monomial-divisor mirror map. Not all
deformations of the complex structure can be represented
by monomial deformations, which are
also referred to as algebraic deformations. We will
however restrict our analysis to those.
We will now describe it for Fermat hypersurfaces of degree $d$, following
\batyrevII.

The toric variety $\IP_{\Delta(\vec w)}$ can be identified with
\eqn\Hfive{
\eqalign{
\IP_{\Delta(\vec w)} &\equiv {\bf H}_d(\vec w) \cr
&=\{[U_0,U_1,U_2,U_3,U_4,U_5]\in\IP^5
|\prod_{i=1}^5 U_i^{w_{i}} = U^d_0 \},
}}
where the variables $X_i$ in eq.\laurentpolynomial~ are related to the
$U_i$ by
\eqn\toricident{
[1,X_1,X_2,X_3,X_4,{1\over \prod_{i=1}^4 X_i^{w_{i}}}] =
[1,{U_1 \over U_0},{U_2 \over U_0},{U_3 \over U_0},
   {U_4 \over U_0},{U_5 \over U_0}].}
Let us consider the mapping
$\phi:\IP^4\(\vec w\)\rightarrow {\bf H}_d(\vec w)$ given by\foot{In toric
geometry this mapping replaces the orbifold
construction for the mirror manifolds described in
\gp.}
\eqn\etalemap{
[z_1,z_2,z_3,z_4,z_5] \mapsto [z_1z_2z_3z_4z_5,
z_1^{d/w_1},z_2^{d/w_2},z_3^{d/w_3},z_4^{d/w_4},z_5^{d/w_5}].}
Integral points in $\Delta(\vec w)$ are mapped to
monomials of the homogeneous coordinates of $\IP^4\(\vec w\)$ by
\eqn\monomialdivisor{
\mu=(\mu_1,\mu_2,\mu_3,\mu_4)\mapsto\phi^*(X^{\mu} U_0)=
{\prod_{i=1}^4 z_i^{\mu_i d/w_i}
\over\left(\prod_{i=1}^5 z_i\right)^{\sum_{i=1}^4\mu_{i}-1}}}
%Since in toric geometry the integral points inside $\Delta^*(\vecw)$
%describe the exceptional divisors which are introduced
%in the process of the resolution of
%the toric variety $\IP_{\Delta(\vec w)}$, the map \monomialdivisor~
%is called the monomial-divisor map.
Note that the Laurent polynomial $f_\Delta$ and the
quasi-homogeneous polynomial $p=\phi^*(U_0 f_\Delta)$
between which the monomial-divisor map acts correspond to a mirror
pair.
The point at the origin of $\Delta$ is always mapped to the symmetric
perturbation $z_1\cdots z_5$ which is always present. It represents
the restriction of the K\"ahler form of the embedding space
$\P_{\Delta^*}$ to the hypersurface.

The situation for CY hypersurfaces in non-Gorenstein $\IP^4\(\vec w\)$'s
was discussed in \hktyI. Here the corresponding polyhedron is
still reflexive but no longer simplicial and the associated
toric variety $\IP_\Delta$ is Gorenstein.

We have already mentioned that Fermat hypersurfaces in weighted
projective space do not intersect with the singular points of the
embedding space. This is no longer generally true for non-Fermat
hypersurfaces. What does however hold is that hypersurfaces
$Z_{f,\Delta}$
in the corresponding (Gorenstein) $\IP_\Delta$ do not intersect singular
points in $\IP_\Delta$. The singular points of the embedding space
correspond to the lattice points in the interior of
faces of $\IP_\Delta$ of codimension one.
The Laurent monomials for these points thus do not correspond to
complex structure deformations and we will in the following always restrict
the sum in \laurentpolynomial~to those lattice points of $\Delta$
which do not lie in the interior of faces of codimension one.

The final point to mention in this section is the computation of
topological triple intersection numbers. They represent the classical
part (field theory limit, large radius limit) of the
$\bra{27}^3\ket$ Yukawa couplings.

Given three divisors the topological triple intersection number
can be computed in terms of an integral of the (by Poincar\'e duality)
associated harmonic forms: $D_i\cdot D_j\cdot D_k=\int_{\hat X}
h_{D_i}\wedge h_{D_j}\wedge h_{D_k}$. In toric geometry their
evaluation reduces to combinatorics. The results are scattered
through the (mathematics) literature \toric
\ref\roancoup{S.-S. Roan:
   {\sl Topological Couplings of Calabi-Yau Orbifolds},
   Max-Planck-Institut Serie No. 22 (1992),
   to appear in {\sl J. of Group Theory in Physics}}
and have been collected in
\hktyI. We will not repeat them here.

%%%%%%%%%%%%%%%%%%%%%%%%%%%%%%%%%%%%%%%%%%%%%%%%%%%%%%%%%%%%%%%%%%%%
\newsec{Periods, Picard-Fuchs Equations and Yukawa Couplings}
%%%%%%%%%%%%%%%%%%%%%%%%%%%%%%%%%%%%%%%%%%%%%%%%%%%%%%%%%%%%%%%%%%%%

Now that we have learned how to construct CY manifolds and
even mirror pairs we can move on towards applying mirror symmetry.
This, as was mentioned before, requires the knowledge of the
$\bra \overline{27}^3\ket$ Yukawa couplings on the mirror manifold
$X^*$ in order to compute the $\bra 27^3\ket$ couplings on $X$.
But in addition we need to know (at least locally) how to map the
complex structure moduli space of $X^*$ to the K\"ahler structure
moduli space of $X$. This is the so-called mirror map. Our discussion will be
restricted to the neighbourhood of the large complex structure limit
of $X^*$ and the large radius limit of $X$ which will be mapped
to each other by the mirror map.

For the purposes of getting the Yukawa couplings and the mirror map
the Picard-Fuchs equations play a crucial role.
So let us turn to them. It is quite easy to explain what they are
but harder to find them explicitly. We will start with the easy part.

We know from the discussion in section two that the dimension of the
third cohomology group is ${\rm dim}(H^3)=b_3=2(h^{2,1}+1)$. Furthermore
we know that the unique holomorphic three form $\Omega$ depends only
on the complex structure. If we take derivatives with
respect to the complex structure moduli, we will get elements
in $H^{3,0}\oplus H^{2,1}\oplus H^{1,2}\oplus H^{0,3}$.
Since $b_3$ is finite, there must be linear relations between
derivatives of $\Omega$ of the
form ${\cal L}\Omega=d\eta$ where $\cal L$ is a differential operator
with moduli dependent coefficients. If we integrate this equation
over an element of the third homology group $H_3$, i.e. over a closed
three cycle, we will get a differential equation
${\cal L}\Pi_i=0$ satisfied by the periods
of $\Omega$. They are defined as
$\Pi_i(a)=\int_{\Gamma_i}\Omega(a)$. $\Gamma_i\in H_3(X,\ZZ)$
and we have made the dependence on the complex structure moduli explicit.
In general we will get a set of coupled linear partial differential equations
for the periods of $\Omega$. These equations are called Picard-Fuchs
equations. In case we have only one complex structure modulus (i.e. $b_3=4$)
one gets just one ordinary (in fact, hypergeometric)
differential equation of order four. For the
general case, i.e. if $b_3\geq 1$, we will describe below how to set
up a complete system of Picard-Fuchs equations.

For a detailed discussion of ordinary differential equations we recommend
the book by Ince
\ref\ince{E.L. Ince, {\sl Ordinary Differential Equations}, Dover
1956}.
For the general case we have profited from the book by Yoshida
\ref\yoshida{M. Yoshida, {\sl Fuchsian Differential Equations},
Braunschweig (1987) Vieweg}.
There exists a vast mathematical literature on PF systems and the
theory of complex moduli spaces; main results are collected in
\ref\griffith{P. A. Griffiths, Bull. Amer. Math. Soc. 76 (1970) 228}.
Two results which are of relevance for our discussion are that
the global monodromy is completely reducible
and that the PF equations have only regular singularities.
The first result enables us to consider only a subset of the
periods by treating only the moduli corresponding to $\tilde h^{2,1}$
out of $h^{2,1}$ moduli. The second result means that the PF equations
are Fuchsian and we can use the theory developped for them.

A systematic, even though generally very tedious procedure
to get the Picard-Fuchs equations for hypersurfaces in
$\IP\(\vec w\)$, is the reduction method due to
Dwork, Katz and Griffiths. As shown in ref.
\ref\griffiths{P. Griffiths, Ann. of Math. 90 (1969) 460}
the periods $\Pi_i(a)$ of the holomorphic three form $\Omega(a)$
can be written as\foot{Again, here and below we only treat the
case of hypersurfaces in a single projective space.
Complete intersections in products of projective spaces
are covered in \hktyII.}
\eqn\periods{\Pi_i(a)=\int_{\Gamma_i}\Omega(a)=\int_{\gamma}
             \int_{\Gamma_i}{\omega\over p(a)}\,,
             \quad i=1,\dots , 2(h^{2,1}+1)\,.}
Here
\eqn\measure{
\omega=\sum_{i=1}^{5}(-1)^{i} w_i z_i dz_1\wedge\dots\wedge
\widehat{dz_i}\wedge\dots\wedge dz_{5}\,;}
$\Gamma_i$ is an element of $H_3(X,\ZZ)$ and
$\gamma$ a small
curve around the hypersurface $p=0$ in the $4$-dimensional embedding space.
$a_i$ are the complex structure moduli, i.e. the coefficients of the
perturbations of the quasi-homogeneous polynomial $p$.
The fact that $\Omega(a)$  as defined above is well behaved is
demonstrated in
\ref\candelas{P. Candelas, \npb298(1988)458}.

The observation that
${\p\over\p z_i}\left({f(z)\over
p^r}\right)\omega$ is exact
if $f(z)$ is homogeneous with degree such that
the whole expression has degree zero,
leads to the partial integration rule, valid under
the integral ($\p_i={\p\over\p z_i}$):
\eqn\partialint{{f\p_i p \over p^{r}}
                ={1\over r-1}{\p_i f\over p^{r-1}}}
In practice one chooses a basis $\{\varphi_k(z)\}$
for the elements of the local
ring ${\cal R}=\IC [z_1,\dots,z_{n+1}]/(\p_i p)$. From the
Poincar\'e polynomial associated to $p$ \lvw~one sees
that there are $(1,\tilde h^{2,1},\tilde h^{2,1},1)$ basis elements
with degrees $(0,d,2d,3d)$ respectively.
The elements of degree $d$ are the perturbing monomials.
One then takes derivatives of the expressions
$\pi_k=\int{\varphi_k(z)\over p^{n+1}}$ ($n={\rm deg}(\varphi_k)/d$)
w.r.t. the moduli.
If one produces an expression such that
the numerator in the integrand is not one of the basis elements,
one relates it, using the equations $\p_i p=...$, to the basis
and uses \partialint. This leads to the so called
Gauss-Manin system of first order differential equations
for the $\pi_k$ which can be rewritten as a system of
partial differential equations for the period.
These are the Picard-Fuchs equations.
In fact, the Picard-Fuchs equations just reflect the structure of the
local ring and expresses the relations between its elements (modulo
the ideal).
It thus depends on the details of the ring how many equations and of which
order comprise a complete system of Picard-Fuchs equations.
To see this, let us consider a model
with $\tilde h^{2,1}=2$, i.e. we have two monomials at degree $d:
\varphi_1$ and $\varphi_2$. Since the dimension of the ring at
degree $2d$ is two, there must be one relation (modulo the ideal)
between the three combinations
$\varphi_1^2,\,\varphi_1\varphi_2,\,\varphi_2^2$.
Multiplying this relation by $\varphi_1$ or $\varphi_2$ leads to two
independent relations at degree three. Since the dimension of the
ring at degree three is one, there must be one further relation at
degree three. The system of Picard-Fuchs equations for models with
$\tilde h^{2,1}=2$ thus consists of one second and one third order equation
\foot{In these considerations we use the fact that the homogeneous
subspace of ${\cal R}$ of degree $nd$ is generated by the elements of
degree $d$ and furthermore that taking a derivative w.r.t. to the
modulus parameter $a_i$ produces one power of the corresponding monomial
$\varphi_i$ in the numerator of the period integral.}.
For models with $\tilde h^{2,1}>2$, a general statement is no longer
possible. For instance, for $\tilde h^{2,1}=3$,
the simplest case is a system of three equations of second order.
The three relations at degree $2d$ then generate all relations at degree
$3d$.  There are however cases where this is not the case and one
has to add extra relations at degree three, leading to third order
equations. Examples of this type are presented in \hktyI.

It is clear from the discussion that the Picard-Fuchs equations
we have obtained only contain those complex structure moduli for which
there exists a monomial perturbation. Also, the method outlined above
applies only to manifolds in projective spaces and not to
manifolds embedded in more general toric varieties.

We will now describe an alternative and often more efficient way to
obtain the Picard-Fuchs equations using the toric data of the hypersurfaces.
The general method has been outlined, in the context of mirror
symmetry, in \batyrevI~ and is based on the generalized hypergeometric
system of Gelf'and, Kapranov and Zelevinsky (GKZ)
\ref\GKZ{I.M.Gel'fand, A.V.Zelevinsky and M.M.Kapranov,
         Func. Anal. Appl. 28 (1989) 12 and
         Adv. Math. 84 (1990) 255}.
We will not decribe it in its generality here, since for our
purposes the following simplified treatment is sufficient.

The way we will proceed is to compute one of the periods, the
so-called fundamental period \bcofhjq\batyrevI~
directly and then set up a system of partial
differential equations satisfied by this period. This system is
the GKZ hypergeometric system. It is not quite yet the PF system
since its solution space is larger than that of the PF system,
which are the periods of the holomorphic three form, of which there are
$2(\tilde h^{2,1}+1)$. (As mentioned before, we are only able to treat
a subset of the $2(h^{2,1}+1)$ periods.)
However the monodromy acts reducibly on the (larger) space of solutions
and the periods are a subset on which it acts irreducibly. It is
often (sometimes?)
easy, starting with the GKZ system,  to find a reduced system of
differential equations which is then the PF system. This will be
explained in the example that we will treat below.

Before showing how to compute the fundamental period and how to
extract the GKZ generalized hypergeometric system, we first have to
discuss the correct choice of coordinates on moduli space. We have
already mentioned that we will only discuss the mirror map in the
neighbourhood of the large complex structure limit. The large complex
structure limit corresponds to the point in moduli space where the
periods have maximal unipotent monodromy
\ref\dmorrison{D. Morrison, {\sl Where is the Large Radius Limit},
preprint IASSNS-HEP-93/68 and {\sl Compactifications of Moduli Spaces
Inspired by Mirror Symmetry}, preprint DUK-M-93-06}.
This in particular means
that the characteristic exponents of the PF equations are maximally
degenerate.
What this means is the following: if we make a power series ansatz
for the solution of the PF equations one gets recursion relations for the
coefficients. However, the condition that
the lowest powers vanish, gives polynomial
equations (of the order of the differential equations)
for the (characteristic) exponents of the lowest order terms
of the power series
\foot{In the case of ordinary differential equations
this polynomial equation is called indicial equations; see e.g. \ince.}.
At the point of
large complex structure the common zeros of these polynomials are all
equal, and in fact, by a suitable moduli
dependent rescaling of the period (this constitutes a choice of gauge, c.f.
below) they can be chosen to be zero. According to the general theory of
Frobenius we then get, in a neighbourhood of large complex structure,
one power series solutions and the other solutions
contain logarithms. We will have more to say about these solutions later.
The problem now consists of finding the
correct variables in which we can write a power series
expansion with these properties.

To find these variables it is necessary to introduce the so-called
lattice of relations. Among the $5+\tilde h^{2,1}$
integer points $\nu_0,\dots,\nu_{\tilde h^{2,1}+5}$ ($\nu_0$ is the origin)
in $\Delta$ which do not lie in the interior of
faces of codimension one, i.e. those
points to which we associate Laurent monomials in
$\laurentpolynomial$,
there are relations of the form $\sum_i l_i v_i=0,\,l_i\in\ZZ$.
The vectors $l$ generate
the $\tilde h^{2,1}$ dimensional lattice of relations.
In this lattice one defines a cone, the so-called Mori cone, whose minimal
set of generators we denote by $l^\a,\,\a=1,\dots,\tilde h^{2,1}$.
(They are also a basis of the lattice of relations.)
This cone is in fact the same as the one mentioned in our brief discussion
about the K\"ahler condition of the resolution of singularities.
We then define the extended vectors
$(\bar l^\a)\equiv(-\sum_i l^\a_i,\{l^\a_i\})\equiv(l^\a_0,\{l^\a_i\})$.
In terms of the parameters appearing in the Laurent
polynomial \laurentpolynomial~the large complex
structure limit is defined to be
the point $u_1=\dots=u_{\tilde h^{2,1}}=0$
in complex structure moduli space with
$u_\a\equiv a^{\bar l^\a}\equiv\prod a_i^{\bar l^\a_i}$.

A systematic method to find the generators of Mori's cone
has been reviewed in \hktyI~where the construction of the upper
convex piecewise linear functions was explained.
An equivalent way\foot{which we have learned
{}from V. Batyrev; see also
\ref\reid{M. Reid, {\sl Decomposition of Toric Morphisms}, in
Progress in
Mathematics 36, M. Artin and J. Tate, eds, Birkh\"auser 1983}.}
is as follows. Consider a particular `triangulation'
(i.e. decomposition into four-simplices with apex at the origin)
of $\Delta$ with lattice points
$\nu_0,\dots,\nu_{\tilde h^{2,1}+5}$. Each four-simplex is then
specified by
four vertices (in addition to the origin $\nu_0$). We then take any pair
of four simplices which have a common three simplex and look for integer
relations $\sum n_i\nu_i=0,\,n_i\in\ZZ$ of the five non-trivial vertices
such that the coefficients of the two vertices which are not common to both
four-simplices are positiv. This provides a set of relations.
There will be $\tilde h^{2,1}$ independent relations in terms of which
the others can be expanded with non-negative integer coefficients. They
constitute the basis of Mori's cone and define the coordinates in the
neighbourhood of large complex structure.

Let us now turn to the computation of the fundamental period. In the
language of toric geometry the period integrals are written as
\foot{This in
fact differs from the period defined before by a factor
of $a_0$, which we have included for later convenience.
This corresponds to a redefinition of $\Omega\to a_0\Omega$.}
\eqn\toricperiod{
\Pi_i(a)=\int_{\Gamma_i}{a_0\over f(a,X)}{dX_1\over X_1}\wedge\cdots
\wedge{dX_4\over X_4}
}
with $\Gamma_i\in H_4((\IC^*)^4\setminus Z_f)$ and $f(a,X)$ the Laurent
polynomial \laurentpolynomial.
We get the fundamental period if we choose the cycle
$\Gamma=\Bigl\{(X_1,\dots,X_4)\in\IC^4\Bigl||X_i|=1\Bigr\}$,
expand the integrand in a power series in $1/a_0$ and evaluate
the integral using the residue formula. Straightforward computation gives
\eqn\fundamentalperiod{
w_0(a)=\sum_{\mu_i\geq0\atop\sum\mu_i\nu_i=0}{(\sum\mu_i)!\over\prod\mu_i!}
{\prod_i a_i^{\mu_i}\over a_0^{\sum\mu_i}}
}
i.e. the sum is over a subset of the lattice of relations. Since the
generators of Mori's cone are a basis of the lattice of relations we
can in fact express the relations to be summed over in the form
$\sum n_\a l^\a$ and sum over the $n_\a$. If we then introduce the
variables $u_\a$, the fundamental period becomes\foot{Here and below
we will use the following notation: for a multi-index
$n=(n_1,\dots,n_N)$
we define $|n|=n_1+\dots+n_N$, $u^n=u_1^{n_1}\cdots u_N^{n_N}$ and also
$n!=\prod n_i!$.}
\eqn\powersol{
w_0(u)=\sum_{\{n_\a\}}{(-\sum_\a l^\a_0 n_\a)!\over\prod_{i>0}
(\sum_\a l^\a_i n_\a)!}\prod_\a u_a^{n_\a}
\equiv\sum_n c(n)u^n
}
where the sum is over those $n_\a$ which leave the arguments
of the factorials non-negative.
It is now straightforward to set up a system of
$\tilde h^{2,1}$ partial differential equations which are satisfied
by $w_0(u)$. Indeed, the coefficients $c(n)$ satisfy recursion relations
of the form $p_\b(n_\a,n_\b+1)c(n_\a,n_\b+1)-
q_\b(n_\a,n_\b)c(n_\a,n_\b)=0$. Here $p_\a$ and $q_\a$ are
polynomials of their respective arguments.
The recursion relations translate to linear differential operators
\eqn\lineardiffop{
{\cal L}_\b=p_\b(\T_\a,\T_\b)-u_\b q_\b(\T_\a,\T_\b)
}
where we have introduced the logarithmic derivatives
$\T_\a=u_\a{d\over d u_\a}$.
The order of the operator ${\cal L}_\b$ equals the sum of the
positive (or negative) components of $\bar l^\b$.
The system of linear differential equations ${\cal L}_\b w(u)=0$
is the generalized hypergeometric system of GKZ.
Since it is related to the polyhedron $\Delta$, it is also
called the $\Delta$-hypergeometric system.

The hypergeometric systems that one gets this way are not generic, but
rather (semi-non)resonant in the language of \GKZ~in which case the
monodromy acts no longer irreducibly. It is in general not straightforward
to extract the PF system from the GKZ system, but in simple cases the
operators ${\cal L}_\a$ factorize ${\cal L}_\a=\ell_\a{\cal D}_\a$
and the ${\cal D}_\a$ form the complete PF system. The general situation,
which might require an extension of the $\Delta$-hypergeometric
system, has been discussed in \hktyI~ and \hktyII.

In any case, when the dust settles, one has a system of PF operators
of the form
\eqn\generalform{
{\cal D}_\a=p_\a(\Theta)+\sum_\b f_{\a\b}(u)q_{\a\b}(\Theta)
}
where $p_\a,q_{\a\b}$ and $f_{\a\b}$ are polynomials with
$f_{\a\b}(0)=0$ and $p_\a$ is homogeneous and $p_\a$ and $q_{\a\b}$ are of the
same degree.
The homogeneity of $p_\a(\Theta)$ follows from
the characterization of the large complex structure
by the requirement that the characteristic
exponents of the PF differential equations
should be maximally degenerate and the gauge choice which
gives a power series solution that starts with a constant.

Note that the terms in \generalform~ of top degree in the $\T$
correspond to
relations between monomials of the same degree modulo the ideal,
whereas the lower order terms correspond to terms in the ideal.
This comment applies if we work with quasi-homogeneous polynomials.

A necessary condition for a
set of period equations to be complete is that there be $2(\tilde h^{2,1}+1)$
degenerate characteristic exponents at $z=0$. The polynomial ring
\eqn\PFring{
{\cal R}=\IC[\Theta_1, \cdots,\Theta_{\tilde h^{2,1}}]/\{p_a(\Theta)\}
}
then has $(1,\tilde h^{2,1},\tilde h^{2,1},1)$ elements at degrees (in $\T$)
$(0,1,2,3)$. This in particular means that the symbols of ${\cal D}_\a$
generate the ideal of symbols
\foot{This is a feature of PF systems
and not generally true for generalized hypergeometric
systems. A counterexample can be found in ref. \yoshida.}.
This observation is important for the
determination of the singular locus of the PF equations. But let us first
explain what is being said \yoshida.

The PF operators ${\cal D}_\a$ define a (left) ideal $I$ in the
ring of differential
operators\foot{Here we work in a neighbourhood $U$
of the origin of a coordinate
system on $\IC^{\tilde h^{2,1}}$.}, i.e. ${\cal D}\in I\Leftrightarrow
{\cal D}w=0$.
The symbol of a partial linear differential
operator ${\cal D}$ in $k$ variables
of order $m$, i.e. ${\cal D}=\sum_{|p|\leq m}
a_p(u)\left({d\over du}\right)^p$, $(|p|=p_1+\cdots+p_k)$ is defined as
$\sigma({\cal D})=\sum_{|p|=m}a_p(u)\xi_1^{p_1}\cdots\xi_k^{p_k}$.
$\xi_1,\dots,\xi_k$ is a coordinate system on the fiber of the
cotangent bundle $T^*U$ at $z=0$.
The ideal of symbols is then $\sigma(I)=\{\sigma({\cal D})|{\cal D}\in I\}$.
The singular locus is $S(I)=\pi({\rm Ch}(I)-U\times\{0\})$
where the characteristic
variety ${\rm Ch}(I)$ is the subvariety in $T^*U$ given by the ideal of
symbols (i.e. given by $\sigma({\cal D})=0,{\cal D}\in I$) and
$\pi$ the projection along the fiber, i.e. setting $\xi=0$. The singular
locus is also the discriminant of the CY hypersurface,
i.e. the locus in moduli space where the manifold fails to be
transverse
\foot{An alternative way to determine the discriminant of
hypersurfaces in
$\IP_\Delta$ was given in \batyrevstraten.}. We will demonstrate the
method introduced here in the example in section seven.

Let us now turn to the discussion of the remaining solutions of
the PF equations in the neighbourhood of the point $u=0$. Due to the fact
that the characteristic exponents are all degenerate, the fundamental
period is the only power series solution. The other ($2\tilde h^{2,1}+1$)
solutions contain logarithms of $u$. We will now show that there are
$\tilde h^{2,1}$ solutions with terms linear in logarithms, the same
number of solutions with parts quadratic and one solution with a part
cubic in logarithms. This corresponds to the grading of the ring
${\cal R}$ (eq.\PFring).

Extending the definition of $x!=\Gamma(x+1)$ to $x\in\IR$,
and that of the coefficients $c(n+\rho)$ in \powersol~for arbitrary values
of $\tilde h^{2,1}$ parameters $\rho_\a$, we define the power series
\eqn\wnullrho{
w_0(u,\rho)=\sum{c(n+\rho)}u^{n+\rho}.
}
Clearly, setting $\rho=0$ gives the fundamental period.
By the method of Frobenius the logarithmic solutions
are obtained by taking linear combinations
of derivatives $D_{\rho}=\sum{1\over(2\pi i)^n}{b_{n}\over n!}
\p_{\rho}^{n}$ of $w_0(u,\rho)$, evaluated at $\rho= 0$.
As $[{\cal D}_\a,\p_{\rho_\a}]=0$ it is then sufficient to check
\eqn\cond{
\left.D_{\rho}\left({\cal D}_\a w_0(u,\rho)\right)
\right|_{\rho=0}=0,\,\,\forall\,\a
}
to establish $D_{\rho} w_0(u,\rho)|_{\rho= 0}$
as a solution. By consideration of the explicit form of the
series \wnullrho~one can show that the conditions for
vanishing of the constant terms in \cond
\eqn\condI{
\left. D_{\rho}(p_\a(\rho)c(0,\rho)u^\rho )
\right|_{\rho=0}=0,\,\,\forall\,\a,
}
are in fact also sufficient.
A moment's thought shows that the following construction of the operators
$D_\rho$ is valid. We consider the ideal ${\cal I}$ in the polynomial ring
$\IC\(\Theta\)$, generated by the $p_\a(\Theta)$.
We endow $\IC\(\Theta\)$ with
a natural vector space structure with the normalized monomials
as orthonormal basis. We can then define ${\cal I}^\perp$ which consists
of the (homogeneous) polynomials orthogonal to the elements in ${\cal I}$.
If we denote a homogeneous element in
${\cal I}^\perp$ by $\sum b_n\t^n$ then the $D_\rho$ are simply
\eqn\identification{
D^{(|n|)}_\rho=\sum{1\over(2\pi i)^{|n|}}{b_n\over n!}\p_\rho^n.
}
Here the sum is over the $n_i$ such that $\sum n_i=|n|$.
The corresponding solutions contain up to $|n|$ powers of logarithms.
Note that since the PF equations are always at least of order two,
the solutions linear in logarithms are
$w_i(u)=D^{(1)}_i w_0(u,\rho)\bigl|_{\rho=0}=
{1\over 2\pi i}{\p\over\p\rho_i}w_0(u,\rho)\bigl|_{\rho=0}\,
(i=1,\dots,\tilde h^{2,1})$.
In section six we will show that as a result of
mirror symmetry, the operators $D^{(3)}$ and $D^{(2)}$ are of the form
$D^{(3)}=-{1\over 3!}{1\over(2\pi i)^3}\sum\kappa^0_{ijk}\p_{\rho_i}
\p_{\rho_j}\p_{\rho_k}$ and $D^{(2)}_i={1\over 2}{1\over(2\pi i)^2}
\sum\kappa^0_{ijk}\p_{\rho_j}\p_{\rho_k}$ with $\kappa^0$ the
topological triple couplings in a basis to be introduced there.

To summarize the discussion above we collect all the solutions to
the PF equations into the period vector:
\eqn\pI{\Pi(z)=\left(\matrix{w_0(u)\cr
D^{(1)}_i w_0(u,\rho)|_{\rho=0}\cr
D^{(2)}_i w_0(u,\rho)|_{\rho=0}\cr
D^{(3)} w_0(u,\rho)|_{\rho=0}}\right).}

The final point we want to discuss in this section is how to compute
the $\bra\overline{27}^3\ket$ Yukawa couplings as
functions of the complex structure moduli. In the next section we
will relate them, via the mirror map, to the $\bra{27}^3\ket$
couplings on the mirror manifold.

We have already given an expression for the $\bra\overline{27}^3\ket$
couplings in eq.\yukawas. There is in fact a more convenient (for what is
going to come) way to write the same expression in terms of bilinears of the
(derivatives of the) periods of $\Omega$
\ref\fs{S. Ferrara and A. Strominger,
in the Proceedings of the Texas A \& M Strings `89 Workshop;
ed. R. Arnowitt et al., World Scientific 1990}
\ref\co{P. Candelas, P. Green and T. H\"ubsch \npb330(1990)49;
P. Candelas and X.C. de la Ossa, \npb355(1991)455}.
For this purpose we introduce an integral basis of $H^3(X,\ZZ)$
with generators $\alpha_i$ and $\beta^j$ ($i,j=0,\dots,h^{2,1}$)
which are dual to a canonical
homology basis $(A^i,B_j)$ of $H_3(X,\ZZ)$ with intersection
numbers $A^i\cdot A^j=B_i\cdot B_j=0,\, A^i\cdot B_j=\delta^i_j$. Then
\eqn\canonicalbasis{
\int_{A^j}\alpha_i=\int_X\alpha_i\wedge\beta^j=
-\int_{B_i}\beta^j=\delta_i^j
}
with all other pairings vanishing. This basis is unique up to
$Sp(2(h^{2,1}+1),\ZZ)$ transformations.

A complex structure on $X$
is now fixed by choosing a particular 3-form as the
holomorphic (3,0) form $\Omega$.
It may be expanded in the above basis of $H^3(X,\Z)$ as
$\Omega=z^i\alpha_i-{\cal F}_i\beta^i$ where
$z^i=\int_{A^i}\Omega,\,{\cal F}_i=\int_{B_i}\Omega$ are periods of
$\Omega$. As shown in
\ref\brgr{R. Bryant and P. Griffiths, {\sl Some observations on the
Infinitesimal Period Relations for Regular Threefolds with Trivial
Canonical Bundle}, in Progress in Mathematics 36, M. Artin and J. Tate, eds;
Birkh\"auser, 1983}
and
\ref\strominger{A. Strominger, \cmp133(1990)163}
the $z^i$ are local complex projective coordinates for the complex
structure moduli space, i.e. we have ${\cal F}_i={\cal F}_i(z)$.
The coordinates $z^i$ are called special projective coordinates. They
are related to the special coordinates of eq.\prepotential, say in a
patch where $z^0\neq0$, i.e. $\l_i={z^i\over z^0}$.
Under a change of complex structure
$\Omega$, which was pure (3,0) to start with,
becomes a mixture of $(3,0)$ and $(2,1)$, i.e.
${\p\over\p z^i}\Omega\in H^{(3,0)}\oplus H^{(2,1)}$. In fact
\tian~
${\p\Omega\over\p z^i}=
k_i\Omega+b_i$ where $b_i\in H^{(2,1)}$ is related to elements in
$H^1(M,T_X)$ via $\Omega$
and $k_i$ is a function of the moduli but
independent of the coordinates of $X$. One immediate consequence is
that $\int\Omega\wedge{\p\Omega\over\p z^i}=0$.
Inserting the expression for $\Omega$ in this equation, one finds
${\cal F}_i={1\over 2}{\p\over\p z^i}(z^j{\cal F}_j)$,
or ${\cal F}_i={\p{\cal F}\over\p z^i}$ with
${\cal F}={1\over 2}z^i{\cal F}_i(z)$, ${\cal F}(\mu z)
=\mu^2 {\cal F}(z)$.
{}From ${\p^2\over\p z^i\p z^k}\Omega\in H^{(3,0)}\oplus
H^{(2,1)}\oplus H^{(1,2)}$ it immediately follows that
also $\int\Omega\wedge{\p^2\over\p z^i\p z^j}\Omega=0$. In fact, this
is already a consequence of the homogeneity of ${\cal F}$.
Finally, ${\p^3\over\p z^i\p z^j\p z^k}\Omega\in H^{(3,0)}\oplus
H^{(2,1)}\oplus H^{(1,2)}\oplus H^{(0,3)}$ and one easily finds
$\int\Omega\wedge{\p^3\over\p z^i\p z^j\p z^k}\Omega
={\p^3\over\p z^i\p z^j\p z^k}{\cal F}
=(z^0)^2{\p^3\over\p \l_i\p \l_j\p \l_k}F$ where ${\cal F}=(z^0)^2 F$
(cf. \prepotential); here $i,j,k=1,\dots,h^{2,1}$.
If one computes the $(0,3)$ part of
$\p^3\Omega$ explicitly \co, one recovers indeed the couplings
$\bar\kappa_{ijk}$ in \yukawas.
{}From the discussion above it also follows that under
a change of coordinates $\l_i\to\tilde \l_i(t)$
the Yukawa couplings transform homogeneously and thus
$\bar\kappa_{ijk}=\int\Omega\wedge\p_i\p_j\p_k\Omega$ holds in any
coordinate system, whereas it can be written as the third derivative
of the prepotential only in special coordinates.
If we redefine $\Omega\to{1\over z_0}\Omega$, the periods are
$(1,\l_i,{\p\over\p \l_i}F,2F-\l_i{\p\over\p \l_i}F)$.
We also note that $K=-\ln\int\Omega\wedge\bar\Omega$, which is easily
shown to be in agreement with
\prepotential, up to a K\"ahler transformation.

We are now ready to link the Yukawa couplings to the PF equations.
In inhomogeneous coordinates $\l_i$ the Yukawa couplings are
\eqn\kappabar{
\bar\kappa_{ijk}=\int\Omega\wedge{\p^3\over\p \l_i\p \l_j\p \l_k}
\Omega=\sum_{l=0}^{\tilde h^{2,1}}(z^l\p_i\p_j\p_k{\cal F}_l
-{\cal F}_l\p_i\p_j\p_k z^l)
}
where $z^l$ and ${\cal F}_l$ are periods of $\Omega$ (in a symplectic
basis, i.e. for the particular choice of cycles
$A^l,\,B_l$ as specified above).

We now define
\eqn\W{
\eqalign{
W^{(k_1,\cdots,k_d)}
&=\sum_l\(z^l\pd_{\l_1}^{k_1}\cdots\pd_{\l_d}^{k_d}{\cal F}_l
               -{\cal F}_l\pd_{\l_1}^{k_1}\cdots\pd_{\l_d}^{k_d}z^l \)
\cr
&:= \sum_l(z^l\pd^{\bf k}{\cal F}_l -{\cal F}_l\pd^{\bf k}z^l ) \;\;.
}}
%We now define ($d=\tilde h^{2,1}$)
%\eqn\W{
%\eqalign{
%W^{(k_1,\cdots,k_d)}
%&=\sum_l\(z^l\pd_{u_1}^{k_1}\cdots\pd_{u_d}^{k_d}{\cal F}_l
%               -{\cal F}_l\pd_{u_1}^{k_1}\cdots\pd_{u_d}^{k_d}z^l \)
%\cr
%&:= \sum_l(z^l\pd^{\bf k}{\cal F}_l -{\cal F}_l\pd^{\bf k}z^l ) \;\;.
%}}
In this notation, $W^{({\bf k})}$ with $\sum k_i=3$
describes the various types of Yukawa couplings
and $W^{({\bf k})}\equiv0$ for $\sum k_i=0,1,2$.
%Here we have used the
%$u_\a$ are the local coordinates of a $\tilde h^{2,1}$ dimensional
%subspace of ${\cal M}_{h^{2,1}}$ as defined before.

If we now write the Picard-Fuchs differential operators in the form
\eqn\picard{
\cD_\a=\sum_{\bf k} f_\a^{({\bf k})}\pd^{\bf k}
\quad ,
}
then we immediately obtain the relation
\eqn\picardrelation{
\sum_{\bf k}f_\a^{({\bf k})}W^{({\bf k})}=0 \quad .
}
Further relations are obtained from operators
$\pd_{\l_\b}\cD_\a$. If the system of PF differential equations is
complete, it is
sufficient for deriving linear relations among the Yukawa couplings
and their derivatives, which can be integrated to give
the Yukawa couplings up to an overall normalization.
In the derivation, we need to use the following relations
which are easily derived
\eqn\yukawarelations{
\eqalign{
W^{(4,0,0,0)}&=2\pd_{\l_1}W^{(3,0,0,0)} \cr
W^{(3,1,0,0)}&=
{3\over2}\pd_{\l_1}W^{(2,1,0,0)}+{1\over2}\pd_{\l_2}W^{(3,0,0,0)} \cr
W^{(2,2,0,0)}&=\pd_{\l_1}W^{(1,2,0,0)}+\pd_{\l_2}W^{(2,1,0,0)} \cr
W^{(2,1,1,0)}&=\pd_{\l_1}W^{(1,1,1,0)}+{1\over2}\pd_{\l_2}W^{(2,0,1,0)}
+{1\over2}\pd_{\l_3}W^{(2,1,0,0)} \cr
W^{(1,1,1,1)}&={1\over2}(\pd_{\l_1}W^{(0,1,1,1)}
+\pd_{\l_2}W^{(1,0,1,1)}
+\pd_{\l_3}W^{(1,1,0,1)}
+\pd_{\l_4}W^{(1,1,1,0)})
\;\; .}}
%\eqn\yukawarelations{
%\eqalign{
%W^{(4,0,0,0)}&=2\pd_{u_1}W^{(3,0,0,0)} \cr
%W^{(3,1,0,0)}&=
%{3\over2}\pd_{u_1}W^{(2,1,0,0)}+{1\over2}\pd_{u_2}W^{(3,0,0,0)} \cr
%W^{(2,2,0,0)}&=\pd_{u_1}W^{(1,2,0,0)}+\pd_{u_2}W^{(2,1,0,0)} \cr
%W^{(2,1,1,0)}&=\pd_{u_1}W^{(1,1,1,0)}+{1\over2}\pd_{u_2}W^{(2,0,1,0)}
%+{1\over2}\pd_{u_3}W^{(2,1,0,0)} \cr
%W^{(1,1,1,1)}&={1\over2}(\pd_{u_1}W^{(0,1,1,1)}
%+\pd_{u_2}W^{(1,0,1,1)}
%+\pd_{u_3}W^{(1,1,0,1)}
%+\pd_{u_4}W^{(1,1,1,0)})
%\;\; .}}
By symmetry the above relations exhaust all possibilities.

We have now described all the calculation that have to be done
on the mirror manifold $\hat X^*$, namely the computation
of the couplings $\bar\kappa_{ijk}(\hat X^*)$\foot{In fact, if we
are only interested in the couplings $\kappa_{ijk}(\hat X)$ we do not
need the $\bar\kappa_{ijk}(\hat X^*)$ explicitly, as will become
clear below.}.
In the next section we show how to go back to the manifold $\hat X$
on which we want to compute the $\bra27^3\ket$ Yukawa
couplings.

%%%%%%%%%%%%%%%%%%%%%%%%%%%%%%%%%%%%%%%%%%%%%%%%%%%%%%%%%%%%%%%%%%%%
\newsec{Mirror Map and Applications of Mirror Symmetry}
%%%%%%%%%%%%%%%%%%%%%%%%%%%%%%%%%%%%%%%%%%%%%%%%%%%%%%%%%%%%%%%%%%%%

The question now arises whether mirror symmetry is merely a
hitherto unknown mathematical curiosity or whether it can also be used
as a practical tool in string theory. The demonstration that this is
indeed the case will be attempted in what follows. We will show how
mirror symmetry allows for the computation of the otherwise difficult
(if not impossible) to get K\"ahler moduli dependence of the
$\bra 27^3\ket$ Yukawa couplings. The methods developped can,
in principle, be applied to any CY hypersurface and incorporates the
dependence of those K\"ahler moduli which correspond to toric divisors.
Given a general model it is however in practice technically rather
cumbersome to actually carry the program through. However, for models
with few moduli it can and has been done successfully. A simple but
non-trivial example will be given in the following section. But before
turning to it, we still need a few ingredients.

In the previous section we have shown how to compute, via the PF equations,
the $\bra\overline{27}^3\ket$ couplings on a Calabi-Yau space given as
a hypersurface of a toric variety. If we now also
want to compute the $\bra 27^3\ket$ couplings, we proceed as follows.
We go to the mirror manifold $\hat X^*$ and compute the
$\bra\overline{27}^3\ket$
couplings there and then use mirror symmetry to go back to $\hat X$.
What this requires is to find the map from the complex structure moduli space
with coordinates $u_\a,\,(\a=1,\dots,\tilde h^{2,1}(\hat X^*))$ to the
K\"ahler structure moduli space on $\hat X$ with coordinates $t_i,\,
(i=1,\dots,\tilde h^{1,1}(\hat X)=\tilde h^{2,1}(\hat X^*))$.
The mirror hypothesis then states that the two Yukawa couplings transform
into each other under this transformation. What one has to take into
account is the transformation properties of the Yukawas under coordinate
transformations: for $t_i\to\tilde t_i(t)$ they transform as
$\kappa_{ijk}(t)={\p\tilde t_l\over\p t_i}{\p\tilde t_m\over\p t_j}
{\p\tilde t_n\over\p t_k}\tilde\kappa_{lmn}(\tilde t(t))$.
Another point to consider is the normalization of the periods and
consequently also of the Yukawas which are quadratic in the periods.
In fact, a change of K\"ahler gauge\foot{This corresponds to a K\"ahler
transformation of the K\"ahler potential
$K=-\log\bigl(\int\Omega\wedge\bar\Omega\bigr)$ of moduli space.}
$\Omega\to f(u)\Omega$ results in an
change of the Yukawa couplings $\bar\kappa_{ijk}\to
f^2(u)\bar\kappa_{ijk}$.
The gauge we choose is such that the fundamental period is one,
i.e. $f(u)=1/w_0(u)$. The $\tilde h^{2,1}$ solutions linear in logarithms
are then
\eqn\newcoord{
t_i(z)={w_i(u)\over w_0(u)}.
}
They serve as the coordinates on the K\"ahler moduli space on $\hat X$
in the neighbourhood of infinite radius which is obtained for
${\rm Im}(t_i)\to\infty$ (recall that ${\rm Im}(t_i)$ are the real
moduli).
Equations \newcoord~ define the mirror map.
This coordinate choice can be identified with the
special coordinates of special geometry\foot{These are the so-called
flat coordinates of the associated topological field theory.}
\strominger
\ref\sg{L. Castellani, R. D'Auria and S. Ferrara, \plb241(1990)57
and \cqg7(1990)1767; S. Ferrara and J. Louis, \plb278(1992)240}.
As discussed in
\ref\cfll{A. Ceresole, R. D'Auria, S. Ferrara,
W. Lerche and J. Louis, \ijmpa8(1993)79},
in these coordinates the Picard-Fuchs differential equations
can be written in the form
\eqn\pfspecial{
\sum_{i=1}^k \p_j \p_p (K_r^{-1})^{li}\p_i\p_r\,\Pi(t)=0,
}
where $K_{ijk}=\p_i\p_j\p_k F$ is derived
{}from the prepotential ${\cal F}=w_0^2 F$ ($\p_i={\p\over \p t_i}$).
This system of fourth order equations can be rewritten as a system of
linear differential equations, the Gauss-Manin system in special coordinates.
The solutions of \pfspecial~
are easily written down in terms of $F$:
$\Pi(t)=(1,t_i,\p_i F,2F-t_i\p_i F)$.
Note that these are the periods in the canonical basis discussed
in section five, after going to inhomogeneous coordinates.
The mirror conjecture now states that $F(t)$ can also be identified with
the prepotential for the K\"ahler structure moduli of the manifold
$X$.

The Yukawa couplings are then
\eqn\kappaI{
\kappa_{ijk}(t)=\p_{t_i}\p_{t_j}\p_{t_k} F(t)={1\over w_0(u(t))^2}
{\p u_\a(t)\over\p t_i}{\p u_\b(t)\over\p t_j}{\p u_\g(t)\over\p t_k}
\bar\kappa_{\a\b\g}(u(t))
%={1\over w_0^2(u)}\p_{z_\a}\p_{z_\b}\p_{z_\g}F(u)
}
Here $\bar\kappa_{\a\b\g}$ are the Yukawa couplings on the mirror
manifold which we showed how to compute in the previous section. In order
to espress the couplings $\kappa_{ijk}$ in terms of the K\"ahler moduli
$t_i$, we have to invert the expressions
$t_i(u)={w_i(u)\over w_0(u)}={1\over 2\pi i}\log(u_i)+O(u)$
which leads to expressions of the form
$u_i=q_i(1+O(q))$. Here we have defined $q_j=e^{2\pi i t_j}$.
This then provides the Yukawa couplings as a power series
in the variables $q_j$.

We now want to compare this with the general form for these Yukawa couplings
given in eq.\instantonsum.
For this we introduce the (multi) degree of the curve $C$, which is defined
as $n_i=\int_C h_i\in\ZZ$ for $h_i\in H^{(1,1)}(X,\ZZ)$. For the
integral
over the K\"ahler form we then get $\int_C J=\sum t_i n_i$.
In terms of the degrees and the variables $q_i$ the Yukawa coupling is
\eqn\kappaII{
\kappa_{ijk}=\kappa^0_{ijk}+\sum_{\{n_i\}}{N(\{n_i\})n_i n_j n_k
\prod_l q_l^{n_l}\over
1-\prod_l q_l^{n_l}}
}
where $N(\{n_i\})$ are integers. In the simplest case of isolated non-singular
rational curves $C$, they give the number of curves at degree $\{n_i\}$.
More generally they have to be interpreted as Euler numbers of a suitably
compactified moduli space of holomorphic maps of degree $\{n_i\}$ from
$\IP^1$ (the genus zero world-sheet) to the CY manifold.

Before turning to the example in the next section, we want to say
a few words about the prepotential.
If we introduce homogeneous coordinates on the K\"ahler structure
moduli space of $\hat X$ via
$t_i=z^i/z^0$,
then the most general ansatz for the prepotential ${\cal
F}(z)=(z^0)^2 F(t)$
which respects homogeneity, is
\eqn\prepot{
\eqalign{{\cal F}&={1\over6}\kappa^0_{ijk}{z^i z^j z^k\over z^0}+{1\over2}
                   a_{ij}z^i z^j+b_i z^i z^0 +
                   {1\over2} c (z^0)^2+{\cal F}_{\rm inst.}\cr
                 &=(z^0)^2\Bigl({1\over6}\kappa^0_{ijk}t_i t_j t_k
                   +{1\over2}a_{ij}t_i t_j+b_i t_i+{1\over2}c
                   +F_{\rm inst.}(t)\Bigr)}}
We have split the prepotential into the classical intersection
part and the instanton part ($F_{\rm inst.}$);
$\kappa^0_{ijk}$ are the classical intersection
numbers. The constants $a_{ij},\, b_i$ and $c$ do not enter the Yukawa
couplings $\kappa_{ijk}(t)=\p_i\p_j\p_k F(t)$. Their real parts are also
irrelevant for the K\"ahler potential. There is a continuous
Peccei-Quinn symmetry $t_i\to t_i+\a_i,\,\a_i$ real
\foot{Under constant shifts of ${\rm Re}(t_i)$ the sigma-model action
\lag~
changes according to $\Delta S\sim\int_\Sigma d^2 z
b_{i\bar\jmath}(\phi)
(\p\phi^i\bar\p\phi^{\bar\jmath}-\bar\p\phi^i\p\phi^{\bar\jmath})
=\int_{\Phi(\Sigma)}b_{i\bar\jmath}d\phi^i\wedge d\phi^{\bar\jmath}
=\int_{\Phi(\Sigma)}b$. For $\Phi(\Sigma)$ topologically
trivial, $b=d a$ and $\Delta S=0$.},
which is broken by instanton corrections to discrete shifts
\ref\ww{X.-G. Wen and E. Witten, \plb166(1986)397}.
Requiring this symmetry in the absence of instanton corrections gives
${\rm Im}(a_{ij})={\rm Im}(b_i)=0$.

{}From the function ${\cal F}$, viewed as the pre-potential for the K\"ahler
structure moduli space on $X$, we construct the vector
$(z^0,z^i,(\p{\cal F}/\p z^i),(\p{\cal F}/\p z^0))\equiv z^0\Pi(t)$ with
\eqn\pIII{
\Pi(t)=\left(\matrix{1\cr
                 t_i\cr
                 {1\over 2}  \kappa^0_{ijk} t_j t_k + a_{ij}t_j +b_i
                 +\p_i(F_{\rm inst.})\cr
                -{1\over 6} \kappa^0_{ijk} t_i t_j t_k+b_i t_i+ c+
                 O(e^{2 \pi i t})}\right)}
The mirror conjecture now says that this is the same as the period
vector \pI.
Comparing the last components of these two vectors, using that
$\log z_j=2\pi i t_j+O(t^2)$ we verify that, up to an overall normalization,
the coefficients of the operator $D_\rho^{(3)}$ are indeed the topological
couplings. We also conclude that the fully instanton corrected couplings
$\kappa_{ijk}$ are given by the concise expressions
\eqn\kappaIII{
\kappa_{ijk}(t)=\p_{t_i}\p_{t_j}{D_k^{(2)}w_0(u(t),\rho)\bigl|_{\rho=0}
\over w_0(u(t))}
}

Let us finally turn to the constants $a_{ij},\,b_i$ and $c$ in the
prepotential. From what has been said we see that they can be expressed
in terms as the expansion coefficient $c(0,\rho)$ in eq.\powersol:
$c=D^{(3)}c(0,\rho)\bigl|_{\rho=0},\,
b_i=D_i^{(2)}c(0,\rho)\bigl|_{\rho=0}$
and $a_{ij}=0$.
One finds
\eqn\rhoderivatives{
\eqalign{\pp{\b}c(0)&=-(l_0^\b+\sum_{i>0}l_i^\b)\Gamma'(1)\equiv 0\cr
         \pp{\b}\pp{\g}c(0)&={\pi^2\over6}
            \Bigl(l_0^\b l_0^\g-\sum_{i>0}l_i^\b l_i^\g\Bigr)\cr
         \pp{\a}\pp{\b}\pp{\g}c(0)&=2\Bigl(l_0^\a l_0^\b l_0^\g+\sum_{i>0}
            l_i^\a l_i^\b l_i^\g\Bigr)\zeta(3)}
}
An interesting observation is now that for all hypersurfaces that we have
treated explicitly, the following relations hold:
\eqn\conjecture{
\eqalign{\chi(\hat X)=\int_{\hat X}c_3&={1\over3}\kappa^0_{\a\b\g}
         \bigl(l_0^\a l_0^\b l_0^\g+\sum_{i>0}l_i^\a l_i^\b l_i^\g\bigr)\cr
         \int_{\hat X}c_2\wedge h_\a&={1\over2}\kappa^0_{\a\b\g}
          \bigl(l_0^\b l_0^\g-\sum_{i>0}l_i^\b l_i^\g\bigr)}
}
For the case of singular hypersurfaces we have no proof of these relations.
For non-singular complete intersections in products of projective spaces
equivalent formulas are derived and proven in \hktyI.
Above results lead to the following expressions for the constants
$b_i$ and $c$:
\eqn\constants{
b_i={1\over24}\int_{\hat X}c_2\wedge h_i\,\qquad
c={1\over (2\pi i)^3}\chi(\hat X)\xi(3)
}
We thus find that $c$, being imaginary, is the only relevant
contribution of $a,b,c$ to the K\"ahler potential.
In fact the value of $c$ reproduce the expected contribution from
the $\sigma$ model loop calculation. Moreover
\constants~reproduces the values for $b$ and $c$ of all
examples where the prepotential is derived by specifying the integral
basis\cdgp\morrison. We therefore conjecture that the prepotential
\prepot~ describes in general the Yukawa
couplings {\it and} the K\"ahler metric
for ${\cal M}_{h^{2,1}}(\hat X^*)$ and therefore by mirror hypothesis
also on ${\cal M}_{h^{1,1}}(\hat X)$ in the region of convergence of the
large complex (K\"ahler) structure expansion.

We now have to comment on the topological couplings
$\kappa^0_{\a\b\g}$.
The indices refer to the coordinates $t_\a$, which, via the
mirror map \newcoord, are related to the $u_\a$. In the large radius
limit we have $t_\a={1\over2\pi i}\log(u_\a)$. The coordinates
$u_\a$ are monomials in the perturbation parameters in the Laurent
polynomial $f_{\Delta}$. These parameters are in one-to-one
correspondence with (exceptional) divisors on $\hat X$. To the coordinates
$u_\a$ we thus have to associate linear combinations of divisors, or,
equivalently, of harmonic $(1,1)$ forms. For Fermat hypersurfaces
this is done in the following way.
If $h_J,h_{D_1},\dots,h_{D_n},\,(n=\tilde h^{2,1}(\hat X^*)-1
=\tilde h^{1,1}(\hat X)-1)$ are
the harmonic $(1,1)$ forms corresponding to the basis $a_0,\dots,a_n$,
the forms corresponding to the basis $u_1,\dots,u_{\tilde h^{2,1}}$ are,
\eqn\basischange{
h_J=h_1\,\quad h_{D_i}=\sum_\a l_{i+5}^\a h_\a
}

%%%%%%%%%%%%%%%%%%%%%%%%%%%%%%%%%%%%%%%%%%%%%%%%%%%%%%%%%%%%%%%%%%%%
\newsec{A Two Moduli Example: Hypersurface in $\IP^4\(2,2,2,1,1\)$}
%%%%%%%%%%%%%%%%%%%%%%%%%%%%%%%%%%%%%%%%%%%%%%%%%%%%%%%%%%%%%%%%%%%%

After all the rather formal discussions, we now want to present an
example for which we choose the degree eight Fermat
CY hypersurface in $\IP^4\(2,2,2,1,1\)$ defined by
\eqn\exampleI{
p_0(z)=z_1^4+z_2^4+z_3^4+z_4^8+z_5^8=0
}
We will end up with the various Yukawa couplings between the to
$27$-plets of $E_6$, in particular their dependence on the K\"ahler
moduli. This will also provide the number of rational curves at
all degrees (instanton numbers).

Being a Fermat model, both $\Delta$ and $\Delta^*$ are simplicial and their
corners can be easily written down using \vertices.
Constructing the remaining lattice points in $\Delta$ and $\Delta^*$
and applying formulas \hodgenumbers~gives $h^{1,1}(\hat X)$=2 and
and $h^{2,1}(\hat X)=86$ and the reversed numbers for the mirror partner.
Note however that $\tilde h^{2,1}(\hat X)=83$. This corresponds to the
fact that we can only incorporate 83 complex structure deformations
as monomial perturbations of the quasi-homogeneous polynomial $p_0$.
In fact, one readily writes down the most general quasi-homogeneous
polynomial of degree eight. It has 105 monomials.
Using the freedom of homogeneous coordinate redefinitions, which provide
22 parameters, we are left with 83 possible monomial deformations
which we might choose to be the degree eight elements of the ring
${\cal R}={\IC\(z_1,\dots,z_5\)\over\{\p_i p_0\}}$.
An easy way to see this is that if we make
infinitesimal homogeneous coordinate transformations
$p_0$ changes (to first order) by terms in the ideal.

We also understand the number of K\"ahler moduli.
The model has a singular $\ZZ_2$ curve $C$ and the exceptional divisor is
$C\times\IP^1$. We thus have $h^{1,1}=2$, one of the forms coming from the
embedding space and the second from the exceptional divisor. Since we have
a singular curve we expect a lattice point on a face of $\Delta$ of dimension
one (i.e. on an edge). This point is
$\nu_6=(-1,-1,-1,0)={1\over 2}(\nu_5+\nu_4)$.
Via the monomial-divisor map it corresponds to the
perturbation $z_4^4 z_5^4$.
The Laurent polynomial for the mirror is thus
\eqn\exampleII{
f_{\Delta}(X)=a_0-\Bigl(a_1 X_1+a_2 X_2+a_3 X_3+a_4 X_4
+{a_5\over X_1^2 X_2^2 X_3^2 X_4}+{a_6\over X_1 X_2 X_3}\Bigr)
}
where we may use the freedom to rescale the variables $X_i$ and the
polynomial to set $a_1=\dots=a_5=1$. Equivalently we can write the
homogeneous polynomial $p(z)=p_0(z)-a_0 z_1\cdots z_5-a_6 z_4^4 z_5^4$.

There are two independent relations between the lattice
points which can be represented by the following two generators of the
lattice of relations: $l^{(1)}=(-4,1,1,1,0,0,1)$ and
$l^{(2)}=(0,0,0,0,1,1,-2)$. They do in fact generate the Mori cone and
thus define the coordinates in the neighbourhood of the large complex
structure point:
$u_1\equiv u={a_1 a_2 a_3 a_6\over a_0^4}$ and
$u_2\equiv v={a_4 a_5\over a_6^2}$.
The fundamental period is
\eqn\exampleIIa{
w_0(u,v)=\sum c(n,m)u^n v^m
}
with
$$
c(n,m)={(4n)!\over(n!)^3(m!)^2(n-2m)!}
$$
The Picard-Fuchs equations
are then found to be ${\cal D}_{1,2}\omega(u,v)=0$ with
\eqn\exampleIII{
\eqalign{{\cal D}_1&=\tu^2(\tu-2\tv)-4u(4\tu+3)(4\tu+2)(4\tu+1)\cr
         {\cal D}_2&=\tv^2-v(2\tv-\tu+1)(2\tv-\tu)}
}
where ${\cal L}_1=\tu{\cal D}_1$ and ${\cal L}_2={\cal D}_2$.

{}From the Picard-Fuchs equations we can read off the symbols
$\sigma({\cal D}_1)$ and $\sigma({\cal D}_2)$ which generate
the ideal of symbols:
\eqn\exampleIV{
\eqalign{\sigma({\cal D}_1)&=u^2\xi_u^2\bigl(u\xi_u(1-4^4 u^2)-2v\xi_v\bigr)\cr
         \sigma({\cal D}_2)&=v^2\xi_v^2-v(2v\xi_v-u\xi_u)^2}
}
To get the discriminant we have to look for simultaneous solutions
of $\sigma({\cal D}_1)=\sigma({\cal D}_2)=0$ other than
$\xi_u=\xi_v=0$, which leads to the characteristic variety ${\rm Ch}(I)$.
Setting $\xi_u=\xi_v=0$ then gives the discriminant.
It is straightforward to verify that ${\rm dis}(\hat X^*)
=\Delta_1\Delta_2\Delta_3\Delta_4$ with
\eqn\exampleV{
\Delta_1=(1-512u+65536u^2-262144 u^2 v)\,,\Delta_2=(1-4v),\,
\Delta_3=u,\,\Delta_4=v
}
being its irreducible components.

The PF equations also determine the Yukawa couplings (up to an overall
multiplicative constant). From the PF equations we derive
three third order equations with operators $\tu{\cal D}_1,\,
\tv{\cal D}_1$ and ${\cal D}_2$ which provide three linear relations
between the four different Yukawa couplings. We can thus express
all of them in terms of one for which we derive two linear
first oder differential equations which can be integrated. The final
result is:
\eqn\exampleVI{
\eqalign{K_{uuu}&=c{1\over\Delta_1\,\Delta_3^3}\,,\quad
K_{uuv}=c{1-256u\over 2\Delta_1\Delta_3^2\Delta_4}\cr
K_{uvv}&=c{512u-1\over \Delta_1\Delta_2\Delta_3\Delta_4}\,,\quad
K_{vvv}=c{1-256u+4v-3072uv\over 2\Delta_1(\Delta_2\Delta_4)^2}}
}
where $c$ is an integration constant which will be fixed below.
These are the $\bra\overline{27}^3\ket$ couplings on the manifold
$\hat X^*$. We now perform the mirror map to compute the
$\bra 27^3\ket$ couplings on $\hat X$. To construct the variables
$t_i={w_i(u)\over w_0(u)}$ we need the solutions
to the PF equations which are linear in logarithms of the variables.
Following our discussion in section five we write
\eqn\exampleVII{
\eqalign{w_i(u)&={1\over 2\pi i}{\p\over\p\rho_i}w_0(u,\rho)\cr
&={1\over 2\pi i}{\p\over\p\rho_i}\sum_{n,m}
{\Gamma(4(n+\rho_1)+1)\over\Gamma^3(n\!+\!\rho_1\!+\!1)
\Gamma^2(m\!+\!\rho_2\!+\!1)
\Gamma(n\!-\!2m\!+\!\rho_1\!-\!2\rho_2\!+\!1)}u^{n\!+\!\rho_1}v^{m\!+\!\rho_2}
\Bigr|_{\rho_1\!=\!\rho_2\!=\!0}\cr
&=w_0(u)\log u_i+\tilde w_i(u)}
}
where
\eqn\exampleVIII{
\tilde w_i(u)={1\over2\pi i}\sum_{m,n}d_i(n,m)u^n v^m
}
with $(\psi=\Gamma'/\Gamma)$
\eqn\exampleIX{
\eqalign{d_1(n,m)&=\{4\psi(4n+1)-3\psi(n+1)-\psi(n-2m+1)\}c(n,m)\cr
         d_2(n,m)&=\{-2\psi(m+1)+2\psi(n-2m+1)\}c(n,m)}
}
We can also write down the remaining solutions of the PF equations.
The ideal ${\cal I}$ is generated by $I_1=\tu^2(\tu-2\tv)$ and
$I_2=\tv^2$, so that a basis of ${\cal I}^\perp$ is
$\{1,\tu,\tv,\tu^2,\tu\tv,
2\tu^3+\tu^2\tv\}$. The elements at degrees zero and one lead, via
\identification,
to the periods $w_0$ and $w_{1,2}$ already given
above. The elements at degrees two and three give the remaining solutions, with
up to two and three powers of logarithms, respectively.

The topological triple couplings, i.e. the infinite radius limit of the
$\bra 27^3\ket$ couplings on $\hat X$, are, using the rules given in
\hktyI~and an obvious notation,  $\kappa^0=8 J^3-8JD^2-16 D^3$.
Here $J$ and $D$ refer to the divisors which correspond to the two lattice
points $\nu_0$ and $\nu_6$, respectively with the associated
moduli parameters $a_0$ and $a_6$. To transform to the divisors whose
associated moduli parameters are $u$ and $v$, we have to take linear
combinations (c.f. \basischange)
and define $D_1=J$ and $D_2={1\over2}(J-D)$. In this basis
the triple intersection numbers are $\kappa^0=8 D_1^3+4 D_1^2 D_2$.
They will be used to normalize the couplings $\bra 27^3\ket$.
Ratios of
the couplings in the latter basis can also be read off from the elements
of ${\cal I}^\perp$: $D_i^{(2)}={1\over2}\kappa^0_{ijk}
\p_{\rho_j}\p_{\rho_k}$ and
$D^{(3)}=-{1\over6}\kappa^0_{ijk}\p_{\rho_i}\p_{\rho_j}\p_{\rho_k}$.

Now we are almost done. What is left to do is to make the coordinate
transformation from the complex structure moduli space of $\hat X^*$
with coordinates $u,v$ to the K\"ahler structure moduli space of
$\hat X$ with coordinates $t_1,t_2$ and to go to the gauge $w_0=1$,
i.e. to divide the Yukawa couplings by $(w_0(z(t)))^2$ and express
them in terms of $q_i=e^{2\pi i t_i}$ (this involves inversion
of power series).
At lowest orders this leads to the expansions
\eqn\exampleX{
\eqalign{\kappa_{111}&=\kappa^0_{111}+N(1,0)\,q_1+(N(1,0)+8
N(2,0))\,q_1^2
                       +N(1,1)\,q_1 q_2+O(q^3)\cr
                     &=8+640\,q_1+80896\,q_1^2+640\,q_1 q_2+O(q^3)\cr
         \kappa_{112}&=\kappa^0_{112}+N(1,1)\,q_1 q_2+O(q^3)
                      =4+640\,q_1 q_2+O(q^3)\cr
         \kappa_{122}&=\kappa^0_{122}+N(1,1)\,q_1 q_2+O(q^3)
                      =640\,q_1 q_2+O(q^3)\cr
         \kappa_{222}&=\kappa^0_{222}+N(0,1)q_2+(N(0,1)+4 N(0,2))\,q_2^2
                       +N(1,1)\,q_1 q_2+O(q^3)\cr
                     &=4\,q_2+4\,q_2^2+640\,q_1 q_2+O(q^3)}
}
{}from which we can read off the numbers of instantons at lowest degrees.
Results at higher degrees can be found in refs.\cdfkm~and \hktyI
\foot{Note that in ref. \hktyI, the degrees
of the rational curve are defined with respect to the basis $h_J,h_D$
as
$n_J=\int_C h_J$ and $n_D=\int_C h_D$ whereas here we define
them with respect to the basis $h_{D_1},h_{D_2}$ as
$(n_1,n_2)=(\int_C h_{D_1},\int_C h_{D_2})$.}.

This completes our discussion of this model, which served as a
demonstration
of the techniques outlined in earlier sections. These techniques have
been applied for models with up to three moduli in \hktyI~ and
\hktyII~ and can easily be extended to even more moduli. The hard part
seems to be to set up the PF equations. It is not always as
easy as in the example above (cf.\hktyI).

%%%%%%%%%%%%%%%%%%%%%%%%%%%%%%%%%%%%%%%%%%%%%%%%%%%%%%%%%%%%%%%%%%%%
\newsec{Conclusions and Outlook}
%%%%%%%%%%%%%%%%%%%%%%%%%%%%%%%%%%%%%%%%%%%%%%%%%%%%%%%%%%%%%%%%%%%%

We have tried to convey an idea of the main concepts necessary to
understand recent developments of Calabi-Yau compactification
of string theory. One of the main advances in the past few years
has been the use of mirror symmetry to compute Yukawa couplings.
The information necessary to get the K\"ahler potential, which is of
course essential in order to normalize the fields and hence the
Yukawa couplings, is also contained in the Picard-Fuchs equations.
This has been done explicitly
for models with $h^{1,1}=1$ in refs.\cdgp~and \morrison, and for
a few models with $h^{1,1}=2$ in \cdfkm.
We have conjectured that, at least in the vicinity of the large radius
limit, we have constructed quite generally
the correct prepotential from which one
can get the K\"ahler metric.
The analysis presented here was restricted to the region in moduli
space close to the large complex structure and large radius. In the references
just cited, this has been extended to the whole moduli space.
Since one expects the internal dimensions,
or, equivalently, the vacuum expectation values of the moduli, to be
of order one (in units of $1/\a'$), one needs expressions for the
Yukawa couplings which are valid in this range.

Even though there has been considerable progress towards the computation
of phenomenologically relevant couplings in strings on Calabi-Yau
manifolds, there is still a lot of work left to do.
The computation of the Yukawa couplings involving the
$E_6$ singlets is one of them. Also, as long as we have no information
on the value of the moduli, the Yukawa couplings are not yet fixed.
A potential for the moduli might be generated by non-perturbative
string effects. At present there is no hope to compute them. Some
information of their possible form can be obtained from studying
functions of the moduli with the correct transformation behaviour
under duality transformations. However, for general CY compactifications
the duality groups are not known, and even if so, one still has to
face the task to construct functions of the moduli which are candidates
for a non-perturbative potential. One would expect that in general
this will not lead to a unique answer.
Also, most of the things said here seem to be restricted to the symmetric
(2,2) theories. An open problem is the treatment of more general
string vacua. We hope to be able, in the not so far future, to report
some progress on some of these issues, may be even at the same occasion.

\vskip1cm

\noindent{\bf Acknowledgement:} S.T. would like the organizers of the
school
for the opportunity to present our
(and others') results and for the very enjoyable week in Helsinki.
We thank S.T. Yau for his collaboration on most of the matters
discussed here.

\listrefs

\bye